\documentclass[twocolumn]{aastex631}

\newcommand\tess{\textit{TESS}}

\shorttitle{HD 21520}
\shortauthors{Nies et al.}
\graphicspath{{./}{figures/}}
\begin{document}

\title{HD 21520 b: a warm sub-Neptune transiting a bright G dwarf}

\author{Molly Nies}
\affiliation{Department of Physics, East Stroudsburg University, East Stroudsburg, PA 18301, USA}

\author[0000-0002-4510-2268]{Ismael Mireles}
\affiliation{Department of Physics and Astronomy, The University of New Mexico, Albuquerque, NM 87106, USA}
% mirelesi@unm.edu

\author{François Bouchy}
\affiliation{Observatoire astronomique de l\'{}Universit\'{e} de Gen\`{e}ve, 51 Chemin des Maillettes, 1290 Versoix, Switzerland}

\author[0000-0003-2313-467X]{Diana Dragomir}
\affiliation{Department of Physics and Astronomy, The University of New Mexico, Albuquerque, NM 87106, USA}
% dragomir@unm.edu

\author[0000-0003-1360-4404]{Belinda A.\ Nicholson}
\affiliation{University of Southern Queensland, Centre for Astrophysics, West Street, Toowoomba, QLD 4350 Australia}
\affiliation{Sub-department of Astrophysics, University of Oxford, Keble Rd, Oxford, United Kingdom, OX13RH}
% belinda.nicholson@unisq.edu.au

\author{Nora L. Eisner}
\affiliation{Center for Computational Astrophysics, Flatiron Institute,
New York, NY 10010, USA}
%neisner@flatironinstitute.org (CHEOPS)

\author[0000-0001-9047-2965]{Sergio G.\ Sousa}
\affiliation{Instituto de Astrof\'{i}sica e Ci\^{e}ncias do Espa\c{c}o, Universidade do Porto, CAUP,
Rua das Estrelas, 4150-762 Porto, Portugal}
%sergio.sousa@astro.up.pt (ESPRESSO spectroscopic parameters)

%karen.collins@cfa.harvard.edu (LCO data analysis)
\author[0000-0001-6588-9574]{Karen A.\ Collins}
\affiliation{Center for Astrophysics \textbar \ Harvard \& Smithsonian, 60 Garden Street, Cambridge, MA 02138, USA}

%%From Steve Howells contributions, Gemini South
\author[0000-0002-2532-2853]{Steve~B.~Howell}
\affil{NASA Ames Research Center, Moffett Field, CA 94035, USA}
% steve.b.howell@nasa.gov

\author{Carl Ziegler}
\affiliation{Department of Physics, Engineering and Astronomy, Stephen F. Austin State University, 1936 North St, Nacogdoches, TX 75962, USA}
% carl.ziegler@sfasu.edu

\author{Coel Hellier}
\affiliation{Astrophysics Group, Keele University, Staffordshire ST5 5BG, U.K.}
% c.hellier@keele.ac.uk

%Now sorting in alphabetical order:
% Minerva Builder
\author[0000-0003-3216-0626]{Brett Addison}
\affil{University of Southern Queensland, Centre for Astrophysics, West Street, Toowoomba, QLD 4350 Australia}
 \affiliation{Swinburne University of Technology, Centre for Astrophysics and Supercomputing, John Street, Hawthorn, VIC 3122, Australia}

% Minerva Architect
\author{Sarah Ballard}
\affiliation{Department of Astronomy, University of Florida, 211 Bryant Space Science Center, Gainesville, FL, 32611, USA}

% Minerva Architect
% REQUESTED NOT TO BE A CO-AUTHOR
% \author{Timothy Bedding}
% \affil{School of Physics, Sydney Institute for Astronomy (SIfA), The University of Sydney, NSW 2006, Australia}

% Minerva Architect
\author{Brendan P. Bowler}
\affil{Department of Astronomy, The University of Texas at Austin, TX 78712, USA}

\author[0000-0001-7124-4094]{C\'{e}sar Brice\~{n}o}
\affiliation{Cerro Tololo Inter-American Observatory, Casilla 603, La Serena, Chile}
% cbriceno@ctio.noao.edu

\author[0000-0002-2361-5812]{Catherine~A.~Clark}
\affil{Jet Propulsion Laboratory, California Institute of Technology, Pasadena, CA 91109 USA}
\affil{NASA Exoplanet Science Institute, IPAC, California Institute of Technology, Pasadena, CA 91125 USA}

%dennis@astrodennis.com  (LCO data analysis)
\author[0000-0003-2239-0567]{Dennis M.\ Conti}
\affiliation{American Association of Variable Star Observers, 185 Alewife Brook Parkway, Suite 410, Cambridge, MA 02138, USA}

%ESPRESSO/CORALIE
\author{Xavier Dumusque}
\affiliation{Observatoire astronomique de l\'{}Universit\'{e} de Gen\`{e}ve, 51 Chemin des Maillettes, 1290 Versoix, Switzerland}
%Xavier.Dumusque@unige.ch 

%CHEOPS
\author[0000-0002-5494-3237]{Billy Edwards}
\affil{SRON, Netherlands Institute for Space Research, Niels Bohrweg 4, NL-2333 CA, Leiden, The Netherlands}
%B.Edwards@sron.nl

\author[0000-0003-2519-6161]{Crystal~L.~Gnilka}
\affil{NASA Ames Research Center, Moffett Field, CA 94035, USA}

%ESPRESSO/CORALIE
%\author{Nolan Grieves}
%\affiliation{Observatoire astronomique de l\'{}Universit\'{e} de Gen\`{e}ve, 51 Chemin des Maillettes, 1290 Versoix, Switzerland}
%Nolan.Grieves@unige.ch

%ESPRESSO/CORALIE
\author{Melissa Hobson}
\affiliation{Observatoire astronomique de l\'{}Universit\'{e} de Gen\`{e}ve, 51 Chemin des Maillettes, 1290 Versoix, Switzerland}
% Melissa.Hobson@unige.ch

% Minerva Architect
\author[0000-0002-1160-7970]{Jonathan Horner}
\affil{University of Southern Queensland, Centre for Astrophysics, West Street, Toowoomba, QLD 4350 Australia}

% Minerva Architect
\author[0000-0002-7084-0529]{Stephen R. Kane}
\affil{Department of Earth and Planetary Sciences, University of California, Riverside, CA 92521, USA}

% Minerva Architect
\author[0000-0003-0497-2651]{John Kielkopf}
\affil{Department of Physics and Astronomy, University of Louisville, Louisville, KY 40292, USA}

%ESPRESSO/CORALIE
\author{Baptiste Lavie}
\affiliation{Observatoire astronomique de l\'{}Universit\'{e} de Gen\`{e}ve, 51 Chemin des Maillettes, 1290 Versoix, Switzerland}
%baptiste.lavie@unige.ch 

\author{Nicholas Law}
\affiliation{Department of Physics and Astronomy, The University of North Carolina at Chapel Hill, Chapel Hill, NC 27599-3255, USA}
% nlaw@unc.edu

%ESPRESSO/CORALIE
\author{Monika Lendl}
\affiliation{Observatoire astronomique de l\'{}Universit\'{e} de Gen\`{e}ve, 51 Chemin des Maillettes, 1290 Versoix, Switzerland}
%Monika.Lendl@unige.ch 

\author[0000-0001-7746-5795]{Colin Littlefield}
\affiliation{Bay Area Environmental Research Institute, Moffett Field, CA 94035, USA}
\affiliation{NASA Ames Research Center, Moffett Field, CA 94035, USA}

% Minerva Architect
\author[0000-0001-5162-1753]{Huigen Liu}
\affil{School of Astronomy and Space Science, Key Laboratory of Modern Astronomy and Astrophysics in Ministry of Education, Nanjing University, Nanjing 210046, Jiangsu, China}

%ESPRESSO/CORALIE
%\author{Christophe Lovis}
%\affiliation{Observatoire astronomique de l\'{}Universit\'{e} de Gen\`{e}ve, 51 Chemin des Maillettes, 1290 Versoix, Switzerland}
%christophe.lovis@unige.ch

\author[0000-0003-3654-1602]{Andrew W. Mann}
\affiliation{Department of Physics and Astronomy, The University of North Carolina at Chapel Hill, Chapel Hill, NC 27599-3255, USA}
% awmann@unc.edu

% Minerva Builder
\author{Matthew W. Mengel}
\affil{University of Southern Queensland, Centre for Astrophysics, West Street, Toowoomba, QLD 4350 Australia}

\author[0000-0002-2702-7700]{Dominic Oddo}
\affiliation{Department of Physics and Astronomy, The University of New Mexico, Albuquerque, NM 87106, USA}
% doddo@unm.edu

% Minerva Builder
\author{Jack Okumura}
\affil{University of Southern Queensland, Centre for Astrophysics, West Street, Toowoomba, QLD 4350 Australia}

%epalle@iac.es     (LCO Key Project time contribution)
\author{Enric Palle}
\affiliation{Instituto de Astrof\'\i sica de Canarias (IAC), 38205 La Laguna, Tenerife, Spain}
\affiliation{Departamento de Astrof\'\i sica, Universidad de La Laguna (ULL), 38206, La Laguna, Tenerife, Spain}

% Minerva Architect
\author[0000-0002-8864-1667]{Peter Plavchan}
\affil{George Mason University, 4400 University Drive MS 3F3, Fairfax, VA 22030, USA}

%ESPRESSO/CORALIE
\author{Angelica Psaridi}
\affiliation{Observatoire de Gen{\`e}ve, Universit{\'e} de Gen{\`e}ve, Chemin Pegasi, 51, 1290 Versoix, Switzerland}
%Angeliki.Psaridi@unige.ch 

\author[0000-0003-4422-2919]{Nuno C.\ Santos}
\affiliation{Instituto de Astrof\'{i}sica e Ci\^{e}ncias do Espa\c{c}o, Universidade do Porto, CAUP,
Rua das Estrelas, 4150-762 Porto, Portugal}
\affiliation{ Departamento de F\'{i}sica e Astronomia, Faculdade de Ci\^{e}ncias, Universidade do Porto, Rua do Campo Alegre, 4169-007 Porto, Portugal}
%nuno.santos@astro.up.pt (ESPRESSO spectroscopic parameters)

%rpschwarz@comcast.net (LCO data analysis)
\author[0000-0001-8227-1020]{Richard P. Schwarz}
\affiliation{Center for Astrophysics \textbar \ Harvard \& Smithsonian, 60 Garden Street, Cambridge, MA 02138, USA}

%shporer@mit.edu  (LCO Key Project PI) (Minerva Architect)
\author[0000-0002-1836-3120]{Avi Shporer} 
\affiliation{Department of Physics and Kavli Institute for Astrophysics and Space Research, Massachusetts Institute of Technology, Cambridge, MA 02139, USA}

% Minerva Architect
% REQUESTED NOT TO BE A CO-AUTHOR
% \author{C.G. Tinney}
% \affil{Exoplanetary Science at UNSW, School of Physics, UNSW Sydney, NSW 2052, Australia}

% Minerva Architect
\author[0000-0001-9957-9304]{Robert A. Wittenmyer}
\affil{University of Southern Queensland, Centre for Astrophysics, West Street, Toowoomba, QLD 4350 Australia}

% Minerva Builder
\author[0000-0001-7294-5386]{Duncan J. Wright}
\affil{University of Southern Queensland, Centre for Astrophysics, West Street, Toowoomba, QLD 4350 Australia}

% Minerva Architect
\author[0000-0003-3491-6394]{Hui Zhang}
\affil{Shanghai Astronomical Observatory, Chinese Academy of Sciences, Shanghai 200030, China}

% POC, GI Office, & MAST contributing author
\author{David Watanabe}
\affiliation{Planetary Discoveries in Fredericksburg, VA 22405, USA}

% POC, GI Office, & MAST contributing author
\author{Jennifer V.~Medina}
\affiliation{Space Telescope Science Institute, 3700 San Martin Drive, Baltimore, MD, 21218, USA}

% POC, GI Office, & MAST contributing author
\author{Joel Villase{\~ n}or}
\affiliation{Department of Physics and Kavli Institute for Astrophysics and Space Research, Massachusetts Institute of Technology, Cambridge, MA 02139, USA}

% SPOC contributing author
\author[0000-0002-8219-9505]{Eric~B.~Ting}
\affiliation{NASA Ames Research Center, Moffett Field, CA 94035, USA}

% TSO/ ExoFOP contributing author
% REQUESTED NOT TO BE A CO-AUTHOR
%\author[0000-0003-0918-7484]{Chelsea~ X.~Huang}
%\affiliation{Department of Physics and Kavli Institute for Astrophysics and Space Research, Massachusetts Institute of Technology, Cambridge, MA 02139, USA}
%\affiliation{Juan Carlos Torres Fellow}
% chelsea.huang@usq.edu.au

% TSO/ ExoFOP contributing author
\author[0000-0002-8035-4778]{Jessie L. Christiansen}
\affiliation{Caltech/IPAC-NASA Exoplanet Science Institute, 770 S. Wilson Avenue, Pasadena, CA 91106, USA}
% christia@ipac.caltech.edu

% TESS architect
\author[0000-0002-4265-047X]{Joshua N.\ Winn}
\affiliation{Department of Astrophysical Sciences, Princeton University, Princeton, NJ 08544, USA}

\author[0000-0002-3481-9052]{Keivan G.\ Stassun}
\affiliation{Department of Physics and Astronomy, Vanderbilt University, Nashville, TN 37235, USA}

% TESS architect
\author[0000-0002-6892-6948]{S.~Seager}
\affiliation{Department of Physics and Kavli Institute for Astrophysics and Space Research, Massachusetts Institute of Technology, Cambridge, MA 02139, USA}
\affiliation{Department of Earth, Atmospheric and Planetary Sciences, Massachusetts Institute of Technology, Cambridge, MA 02139, USA}
\affiliation{Department of Aeronautics and Astronautics, MIT, 77 Massachusetts Avenue, Cambridge, MA 02139, USA}
% TESS architect
\author[0000-0001-9911-7388]{David~W.~Latham}
\affiliation{Harvard-Smithsonian Center for Astrophysics, 60 Garden St, Cambridge, MA 02138, USA}

% TESS architect
\author[0000-0003-2058-6662]{George R. Ricker}
\affiliation{MIT Kavli Institute for Astrophysics and Space Research \& MIT Physics Department}

%\collaboration{6}{(AAS Journals Data Editors)}



\begin{abstract}
We report the discovery and validation of HD 21520 b, a transiting planet found with \tess\ and orbiting a bright G dwarf (V=9.2, $T_{eff} = 5871 \pm 62$\, K, $R_{\star} = 1.04\pm 0.02\, R_{\odot}$). HD 21520 b was originally alerted as a system (TOI-4320) consisting of two planet candidates with periods of 703.6 and 46.4 days. However, our analysis supports instead a single-planet system with an orbital period of $25.1292\pm0.0001$ days and radius of $2.70 \pm 0.09\, R_{\oplus}$. Three full transits in sectors 4, 30 and 31 match this period and have transit depths and durations in agreement with each other, as does a partial transit in sector 3. We also observe transits using CHEOPS and LCOGT. SOAR and Gemini high-resolution imaging do not indicate the presence of any nearby companions, and \textsc{Minerva}-Australis and CORALIE radial velocities rule out an on-target spectroscopic binary. Additionally, we use ESPRESSO radial velocities to obtain a tentative mass measurement of $7.9^{+3.2}_{-3.0}\, M_{\oplus}$, with a 3-$\sigma$ upper limit of 17.7 $M_{\oplus}$. Due to the bright nature of its host and likely significant gas envelope of the planet, HD 21520 b is a promising candidate for further mass measurements and for atmospheric characterization. 
\end{abstract}

%\keywords{Extrasolar gaseous planets (2712) -- }


\section{Introduction} \label{sec:intro}
To date, there are thousands of confirmed exoplanets, of which a few hundred (and counting) have been discovered by the Transiting Exoplanet Survey Satellite (\tess; \citealt{Ricker15}). With sky coverage 400 times that of Kepler \citep{Koch10}, the goal of \tess\ is to identify the nearest and brightest transiting systems, namely the ones best suited for confirmation and characterization, in order to enable further understanding of planet and system formation. A desired group of planets to discover are those with long periods as their properties are less understood than for shorter period planets. However, the more planets that are discovered that add to the population, the better we can understand planet formation. 

There is a ``radius valley" that presently exists in the known shorter period ($<$ 100 days) exoplanet population corresponding to fewer exoplanets at $\sim1.8 R_{\oplus}$ Earth radii \citep{Ful17}. The super-Earth planets that are on the lower side ($< 1.8 R_{\oplus}$) of the valley are thought to be rocky while the sub-Neptunes on the higher end of the valley ($> 1.8 R_{\oplus}$) also require volatiles, such as H and He, and/or water envelopes. There are multiple theories regarding the interior structure of sub-Neptune sized planets, as well as multiple theories regarding the origin of the radius valley, namely photoevaporation \citep{Owe17} and core-powered mass loss \citep{Gup19}. The discovery of new longer period planets allows for another dimension in which to investigate the radius valley, and characterize the atmospheres of less irradiated sub-Neptunes to gain a better understanding of their formation. These planets above the radius valley are predicted to have significant H/He envelopes that were not stripped through photoevaporation or core-powered mass loss. However, they could also be ``water worlds", with a significant fraction of water in their interiors and atmospheres \citep{2019PNAS..116.9723Z}.

In this paper, we present the validation of system HD 21520, a sun-like G star with a sub-Neptune planet orbiting at 25.13 day period discovered by \tess. \tess\ has already discovered dozens of sub-Neptune ($R_p < 4 R_{\oplus}$) planets with periods $>$ 20 days, but there are only 4 other planets within this category with host stars both this bright ($V_{mag} = 9.1$) and having an effective temperature close to that of our Sun (HD 63433 c, \citealt{Mann20}; HD 191939 c and d, \citealt{OrellMiquel23}; HD 22946 d \citealt{Garai23}).

We detail the \tess\ photometric observations, CHEOPS photometric observations, LCOGT photometric observations, SOAR high resolution speckle imaging, Gemini South high-resolution images, and \textsc{Minerva}-Australis radial velocity observations, ESPRESSO radial velocity observations, and CORALIE radial velocity observations in Section \ref{sec:observations}. In section \ref{sec:analysis} we analyze these observations to get the stellar and planetary information, as well as statistically validate and eliminate false positive scenarios. In Section \ref{sec:discussion} we discuss the importance of the planet, and in Section \ref{sec:summary} we summarize our findings.

\section{Observations} \label{sec:observations}

\subsection{\tess\ Photometry}\label{subsec:tess}
HD 21520 b was first observed by \tess\ in sectors 3 and 4 of the primary mission at 2 minute cadence, and in Full Frame Images (FFIs) at 30 minute cadence (UT 2018 September 21 to UT 2018 November 14). In the extended mission, HD 21520 was observed again in consecutive sectors 30 and 31 in FFIs at 10 minute cadence, and at 2 minute cadence (UT 2020 September 23 to UT 2020 November 18). The TESS Science Processing Operations Center Pipeline \citep[SPOC;][]{2016SPIE.9913E..3EJ} at NASA Ames Research Center calibrated the FFIS and processed the 2-minute data, producing two light curves per sector called Simple Aperture Photometry (SAP) and Presearch Data Conditioning Simple Aperture Photometry \citep[PDCSAP;][]{2012PASP..124.1000S, 2012PASP..124..985S, 2014PASP..126..100S}, the latter of which is corrected for instrumental signatures, screened for outliers, and corrected for crowding effects. The TESS-SPOC pipeline \citep{2020RNAAS...4..201C} extracted photometry from the SPOC-calibrated FFIs.

HD 21520 b was initially found as a single-transit planet candidate in \tess\ sector 4 by the Planet Hunters \tess\ citizen science project \citep{2021MNRAS.501.4669E}. From their fit to the single transit, they found a period of 26.83$^{+56.14}_{-9.46}$ days, which is in agreement with the true period found in this paper. The transit signature of HD 21520 b was subsequently detected by the SPOC in a multi-sector transit search \citep{2002ApJ...575..493J,2010SPIE.7740E..0DJ,2020TPSkdph} of sectors 3 through 31 on 27 July 2021, which identified two separate threshold crossing events (TCEs) at periods of 703.6 and 46.4 days that were due to the same planet. The signatures passed all the Data Validation diagnostic tests \citep{Twicken:DVdiagnostics2018,Li:DVmodelFit2019} and were promoted to TESS Object of Interest \citep{2021ApJS..254...39G} status on 6 July 2021 as a system containing two planet candidates. This happened because there are three full transits in sectors 4, 30 and 31, and the partial transit in sector 3 was flagged as bad in the SAP light curves. As there are over 700 days between sectors 4 and 30, this resulted in the incorrect identification of two planet candidates with the periods noted above. From the investigation of the PDCSAP lightcurves in sector 3, it is clear that at the end of sector 3 is a partial transit which supports the period of a single planet candidate at 25.13 days. In Section \ref{subsec:planetary_params}, we discuss the fitting of all 4 transits and the planetary parameters that result.

\begin{figure*}[!h]
    \includegraphics[width=\textwidth]{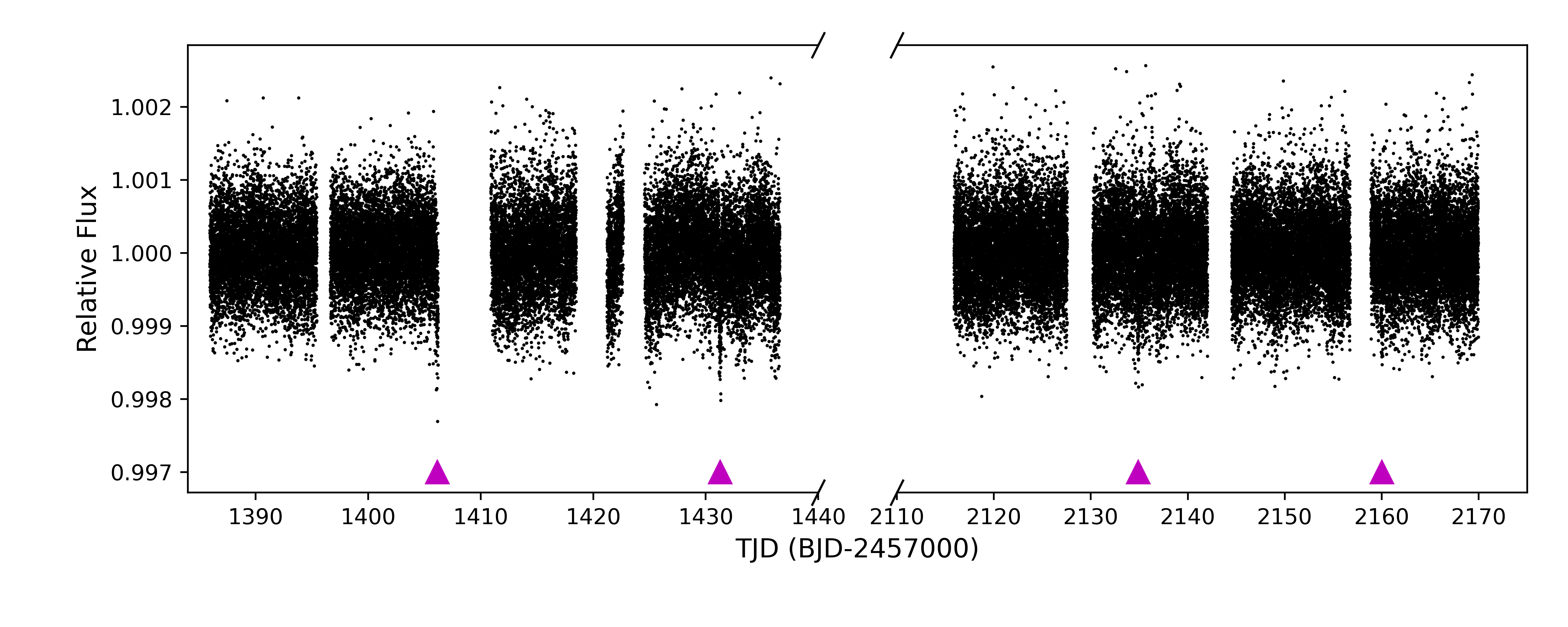}
    \includegraphics[width=\textwidth]{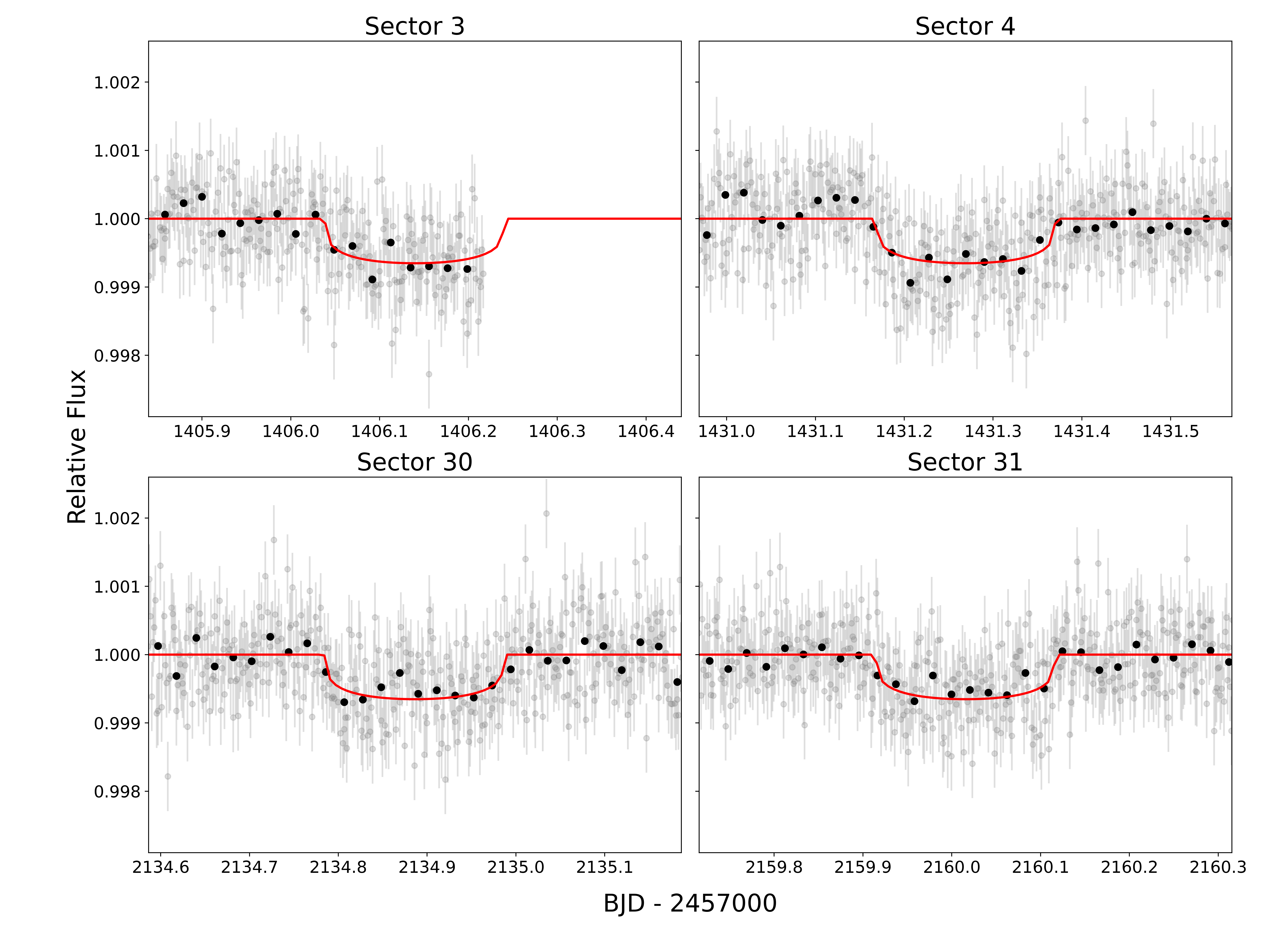}
    \caption{\textit{Top:} Full \tess\, detrended PDCSAP lightcurve of HD 21520 showing the four transits detected (magenta). \textit{Bottom:} Zoom-in on the individual transits showing one partial transit in sector 3 and three full transits in sectors 5, 30, and 31. Gray points are the 2-minute cadence data while black points are the data binned to 30 minutes and the red line is the best-fit light curve.}
    \label{fig:full_lc}
\end{figure*}

In Figure \ref{fig:full_lc} is the full PDCSAP lightcurve, and the individual transits; note the partial transit in sector 3. Fitting of the lightcurve and finding planetary parameters is detailed in Section \ref{subsec:planetary_params}.

\begin{figure*}
\includegraphics[width=\textwidth]{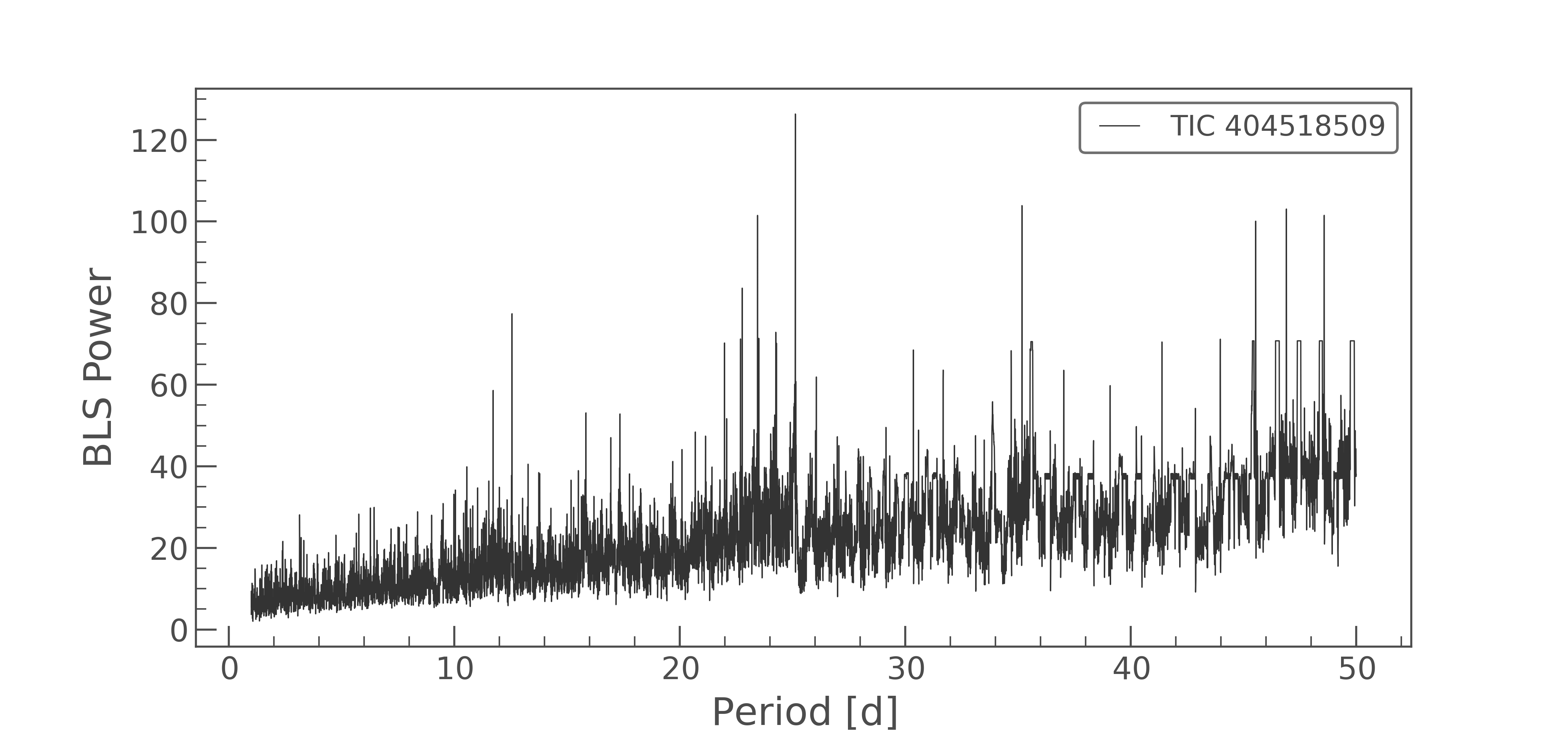}
  \caption{BLS periodogram containing periods less than 50 days, using \tess\ SPOC light curve data. The highest peak is shown at 25.13 days, meaning this periodogram shows agreement with the fitted model period from \texttt{allesfitter}.}
\label{fig:periodogram}
\end{figure*}

A supporting method to confirm the period of HD 21520 b is by generating a Box Least Squares (BLS) periodogram \citep{Kovacs02} and analyzing significant signals. By first using the full \tess\ lightcurve, we normalize, flatten, and remove outliers from the full lightcurve. Then, using \texttt{lightkurve}'s implementation of the BLS method in \texttt{astropy}, we plot the likelihood of the BLS in a period space less than 50 days. The reason to only analyze in this region is due to the large gap of about 700 days between sectors 4 and 30. Including too long of a period range would allow periods that would be aliases arising due to the gaps between sectors, and thus these periods could not be verified with the \tess\ data alone. In Figure \ref{fig:periodogram}, we show that the period found in the \texttt{allesfitter} fit is the highest peak in the periodogram, however there are also other periods with significant power. After investigating each of the other periods with comparable power, we have determined that none of them indicate a second transiting planet candidate. They primarily correspond to harmonics of the true period and in some cases aliases due to gaps in the TESS times series. 

\subsection{CHEOPS Photometry}\label{sec:cheops}

\begin{figure}
\includegraphics[width=\columnwidth]{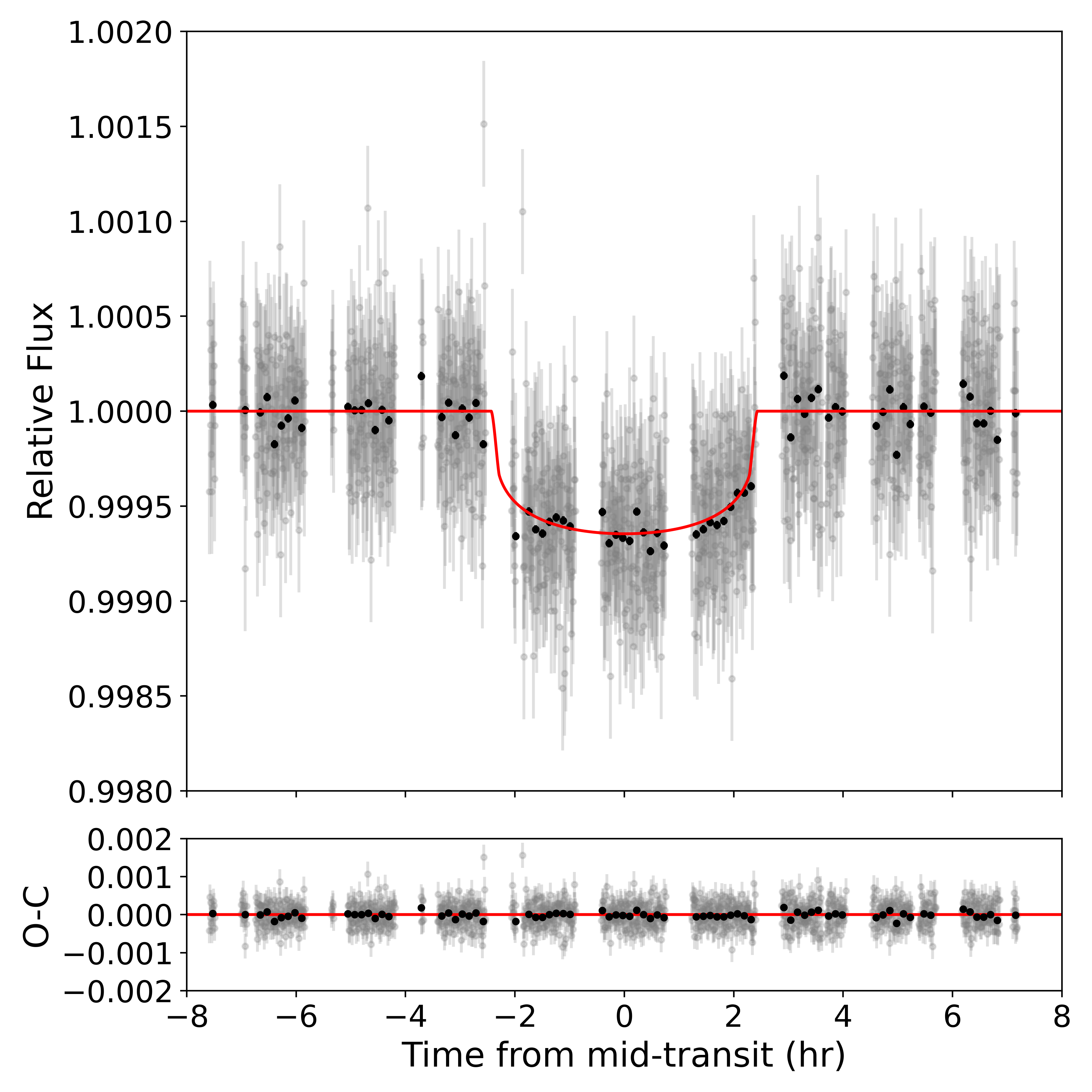}
  \caption{Detrended binned (black) and unbinned (gray) \textit{CHEOPS} photometry showing a clear transit detection. The best-fit light-curve model is shown in red.}
\label{fig:cheops_phot}
\end{figure}

We used the CHaracterising ExOPlanets Satellite \citep[\textit{CHEOPS}; ][]{Benz2021cheops} to observe an additional transit event of HD 21520 b (PN: AO3-27; PI: N. Eisner). The small class ESA mission, which is dedicated to aiding the characterization of known exoplanets, is a 0.32-m Ritchey–Chretien telescope observing at a wavelength range of 0.4 $\mu$m -- 1.1 $\mu$m. The \textit{CHEOPS} observations of HD 21520~b, consisting of 1133 brightness measurements, were obtained between 2023 October 22 18:01 and 2023 October 23 08:47 UTC (9 orbits over $\sim$ 14.75 hours), using exposure times of 30 seconds. The raw data were automatically processed by the \textit{CHEOPS} Data Reduction Pipeline \citep[DRP version 14.1.3; ][]{Hoyer2020cheops}. In brief, the DRP performs the instrumental calibrations (event flagging, bias and gain corrections, linearisation, dark current, and flat field corrections), environmental corrections (cosmic rays, background, and smearing), and performs aperture photometry using four different aperture sizes. While three of the apertures have a fixed radius (RIFN=22.5 pix, DEFAULT=25 pix and RSUP=30 pix), the size of fourth is determined independently for each target, based on the level of contamination in the field of view. In this work, we use the light curves obtained with the DEFAULT=25 pix aperture, as we found this to produce a light curves with the least dispersion when compared to the other photometric apertures, and clearly detect an on-time transit (Figure \ref{fig:cheops_phot}).

\subsection{LCOGT Follow-up}
The \textit{TESS} pixel scale is $\sim 21\arcsec$ pixel$^{-1}$ and photometric apertures typically extend out to roughly 1 arcminute, generally causing multiple stars to blend in the \textit{TESS} aperture. To attempt to determine the true source of the HD 21520 SPOC detections in the \textit{TESS} data, we conducted ground-based photometric follow-up observations of the field around HD 21520 as part of the {\tt TESS} Follow-up Observing Program\footnote{https://tess.mit.edu/followup} Sub Group 1 \citep[TFOP;][]{collins:2019}. 

We observed HD 21520 three times in Pan-STARRS $z$-short band using the Las Cumbres Observatory Global Telescope \citep[LCOGT;][]{Brown:2013} 1.0\,m network nodes at Cerro Tololo Inter-American Observatory (CTIO) on UTC 2021 September 27, South Africa Astronomical Observatory (SAAO) on UTC 2021 November 12, and again from CTIO on UTC 2022 November 05. The 1\,m telescopes are equipped with $4096\times4096$ SINISTRO cameras having an image scale of $0\farcs389$ per pixel, resulting in a $26\arcmin\times26\arcmin$ field of view. The images were calibrated by the standard LCOGT {\tt BANZAI} pipeline \citep{McCully:2018}, and photometric data were extracted using {\tt AstroImageJ} \citep{Collins:2017}. 

The two observations in 2021 were scheduled according to the nominal public SPOC ephemeris for HD 21520.02 (reference epoch T0 = 1417.466332\,BTJD and orbital period P = 46.409373\,d). The UTC 2021 September 27 observation covered an ingress with 75 minutes of baseline and 170 minutes of in-transit coverage under photometric skies. The telescope lost precise guiding after 130 minutes of in-transit coverage, resulting in an a $\sim +2500$\,ppm offset in the lightcurve. We have removed 40 minutes of affected data from our analysis to avoid the need for detrending a short segment of data at the end of the lightcurve. The UTC 2021 November 12 observation covered an egress with 45 minutes of baseline and 180 minutes of in-transit coverage, but suffered from moderate excess sky transparency losses during the final 90 minutes of observations. The ephemeris uncertainty at these epochs was $\sim 15$ minutes, resulting in $\pm5\sigma$ and $\pm3\sigma$ time coverage of ingress and egress, respectively, and after combining and phase-folding, full-transit coverage.

The 2022 November 05 observation was scheduled according to an orbital period alias (1/28) of the SPOC nominal public SPOC ephemeris for HD 21520.02 (resulting in reference epoch T0 = 1431.2688\,BTJD and orbital period P = 25.1297\,d).

For all three observations, we extracted lightcurves of the target star and all 6 known Gaia DR3 and TICv8 neighboring stars within $2\farcm5$ of HD 21520 that are bright enough in \textit{TESS} band to produce the \textit{TESS} detection (allowing for an extra 0.5 magnitudes fainter in \textit{TESS} band to attempt to accommodate uncertainties). We calculate the RMS of each of the 6 nearby star lightcurves (binned in 5 minute bins) and find that the LCOGT lightcurve RMS values are smaller by at least a factor of 3 compared to the expected NEB depth in each respective star. We then visually inspected the neighboring star lightcurves to ensure no obvious deep eclipse-like signal. We therefore rule out a nearby eclipsing binary (NEB) at both ephemerides checked. Our nearby star follow-up lightcurves are available on the {\tt EXOFOP-TESS} website\footnote{\href{https://exofop.ipac.caltech.edu/tess/target.php?id=404518509}{https://exofop.ipac.caltech.edu/tess/target.php?id=404518509}}.

\begin{figure}
\includegraphics[width=\columnwidth]{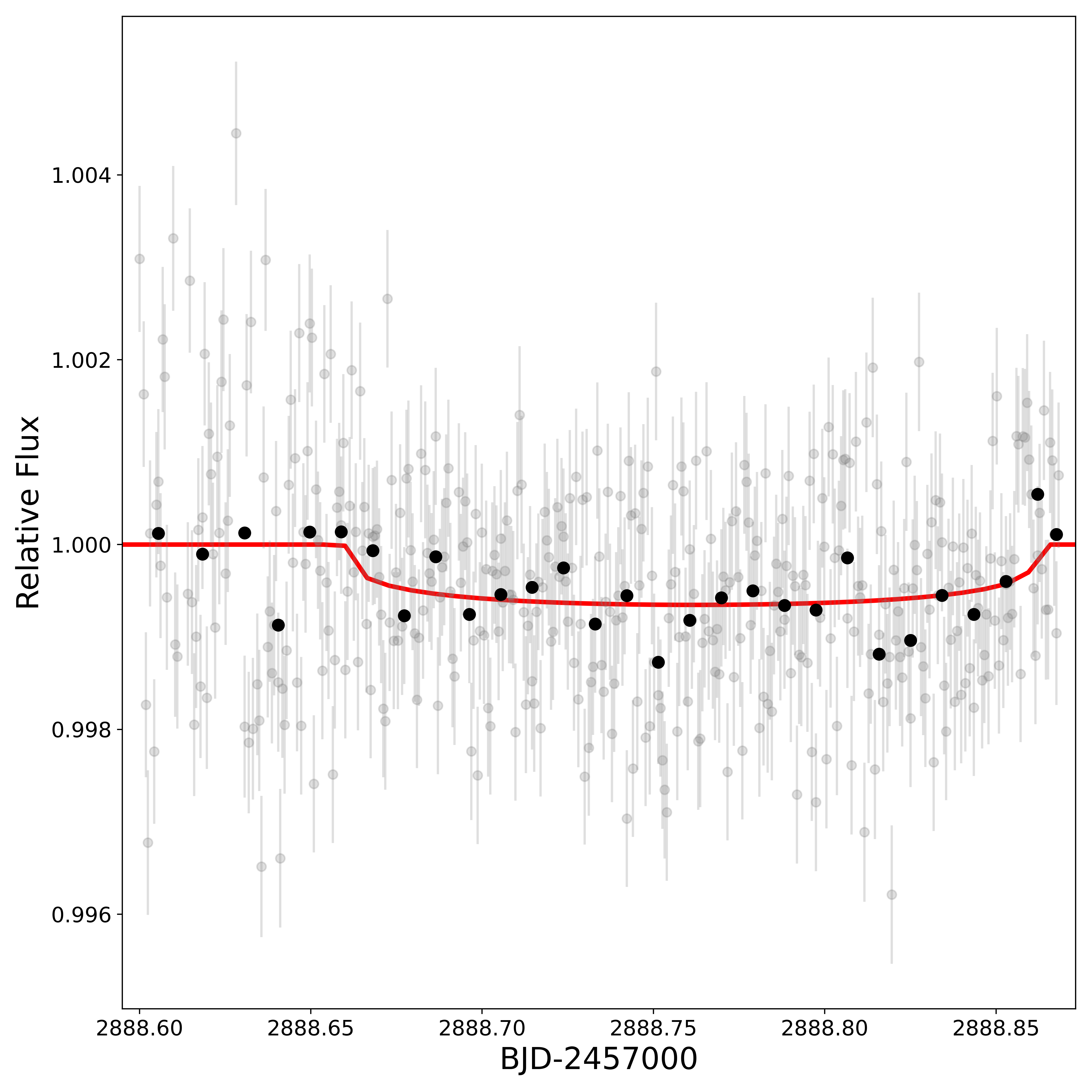}
  \caption{Unbinned (gray) and binned (black) ground based photometry from LCO showing a possible transit detection. The best-fit light-curve model from the joint TESS+CHEOPS+RV fit is shown in red.}
\label{fig:lco_phot}
\end{figure}

The UTC 2021 September 27 lightcurve of the target star HD 21520 has a 10 minute binned scatter of 280 ppm RMS after removing the final 40 minutes of data and shows no evidence of the 430 ppm ingress predicted by the SPOC HD 21520.02 ephemeris. Due to the excess sky transparency losses during part of the UTC 2021 November 12 observation and short baseline coverage, the on-target lightcurve is not sensitive enough for a 400-600 ppm transit egress detection. However, our full transit observation on UTC 2022 November 05 at the revised SPOC HD 21520.01 ephemeris shows a tentative $\sim 4.6$\,hour-long $\sim 700$\,ppm transit detection (see Figure \ref{fig:lco_phot}), which is consistent with the $\sim 4.9$\,hour-long $\sim 630$\,ppm transit detection from \tess.

In summary, we rule out an NEB as the source of the SPOC HD 21520 event detections, probably rule out the event on-target at the HD 21520.02 ephemeris, and likely confirm that an event occurs on-target (relative to known Gaia DR3 and TICv8 stars) at the HD 21520.01 alias having period P = 25.1297\,d.

\subsection{High Resolution Speckle Imaging from SOAR\label{subsec:SOAR}}

\begin{figure}
\includegraphics[width=\columnwidth]{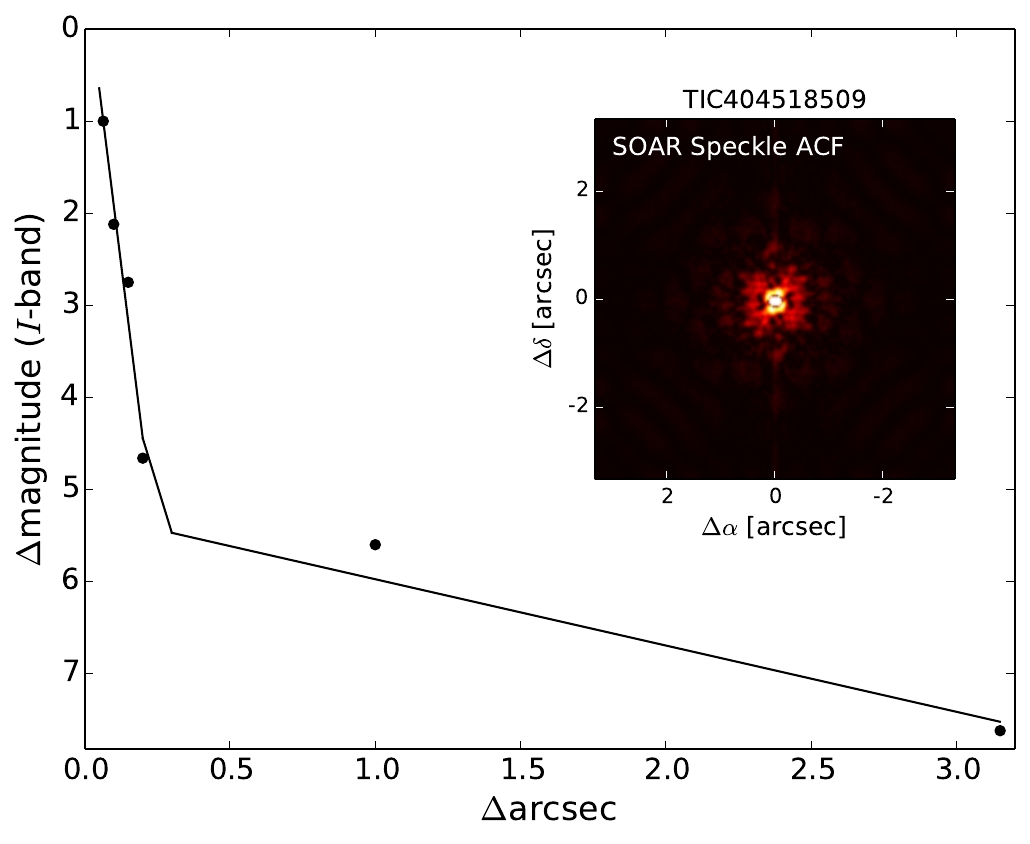}
  \caption{High resolution speckle imaging from SOAR, in addition to the contrast curve.}
\label{fig:soar}
\end{figure}

High-angular resolution imaging is needed to search for nearby sources that can contaminate the TESS photometry, resulting in an underestimated planetary radius, or be the source of astrophysical false positives, such as background eclipsing binaries. We searched for stellar companions to HD 21520 with speckle imaging on the 4.1-m Southern Astrophysical Research (SOAR) telescope \citep{tokovinin2018} on 10 October 2021 UT, observing in Cousins I-band, a similar visible bandpass as TESS. This observation was sensitive to a 5.6-magnitude fainter star at an angular distance of 1 arcsec from the target. More details of the observations within the SOAR TESS survey are available in \citet{ziegler2020}. The 5$\sigma$ detection sensitivity and speckle auto-correlation functions from the observations are shown in Figure~\ref{fig:soar}. No nearby stars were detected within 3\arcsec of HD 21520 in the SOAR observations.

\subsection{Gemini South High-Resolution Imaging}
Close stellar companions (bound or line of sight) can confound exoplanet discoveries in a number of ways.  The detected transit signal might be a false positive due to a background eclipsing binary and even real planet discoveries will yield incorrect stellar and exoplanet parameters if a close companion exists and is unaccounted for.

HD 21520 (TOI-4320) was observed on 2022 October 7 UT using the Zorro speckle instrument on the Gemini South 8-m telescope (Scott et al. 2021, Howell and Furlan 2022).  Zorro provides simultaneous speckle imaging in two bands (562 nm and 832 nm) with output data products including a reconstructed image with robust contrast limits on companion detections.

\begin{figure}
\includegraphics[width=\columnwidth]{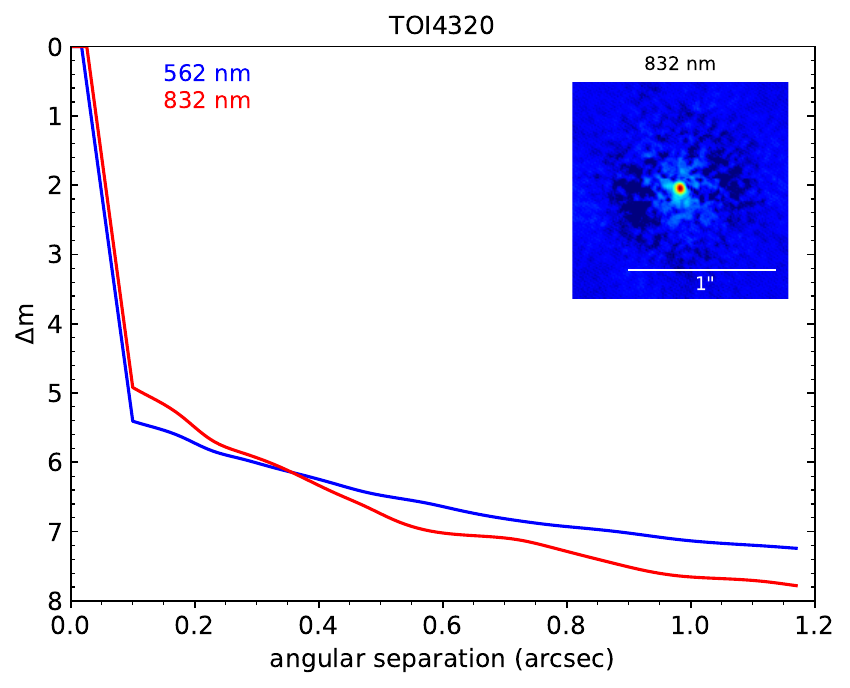}
  \caption{We show the 5$\sigma$ speckle imaging contrast curves in both filters as a function of the angular separation from the diffraction limits out to 1.2 arcsec, the end of speckle coherence. The inset shows the reconstructed 832 nm image with a 1 arcsec scale bar. The star, HD 21520, was found to have no close companions to within the contrast levels achieved.}
\label{fig:gemini}
\end{figure}

Three sets of 1000 $\times$ 0.06 second images were obtained and processed in our standard reduction pipeline (see Howell et al. 2011). From our analysis, HD 21520 was imaged as a single star to within the angular and magnitude contrast levels achieved. Figure \ref{fig:gemini} shows our final 5-sigma contrast curves and the 832 nm reconstructed speckle image. We find that HD 21520 has no close companion brighter than 5-8 magnitudes below that of the target star from the angular limits of the 8-m telescope diffraction limit (20 mas) out to 1.2”. At the distance of HD 21520 (d= 79 pc) these angular limits correspond to spatial limits of 1.6  to 95 AU. Additionally, Gaia DR3 astrometry is consistent with a single star model, as HD 21520 has a RUWE value of 1.03 \citep{2021A&A...649A...2L}.

\subsection{\textsc{Minerva}-Australis Radial Velocities\label{subsec:rvs}}
Radial velocity observations were taken with the \textsc{Minerva}-Australis facility, located at Mt. Kent Observatory in Southern Queensland, Australia \citep{2018arXiv180609282W,2019PASP..131k5003A,2021MNRAS.502.3704A}. \textsc{Minerva}-Australis consists of an array of four independently operated, robotic 0.7m CDK700 telescopes, which simultaneously feed stellar light via fiber optic cables to a single KiwiSpec R4-100 high-resolution (R = 80,000) spectrograph (Barnes et al. 2012) with wavelength coverage from 480 to 620 nm. HD 21520 was observed 40 times between 25th December 2020 and 29th September 2022 with at least two telescopes, and up to four for any one exposure. Radial velocities were computed for each individual telescope's spectrum by cross-correlating the observed spectrum with a master spectrum of the star created by calculating a median of all observations. 

\subsection{CORALIE Radial Velocities\label{subsec:coralie}}

\begin{figure}[!h]
    \centering
    \includegraphics[width=\columnwidth]{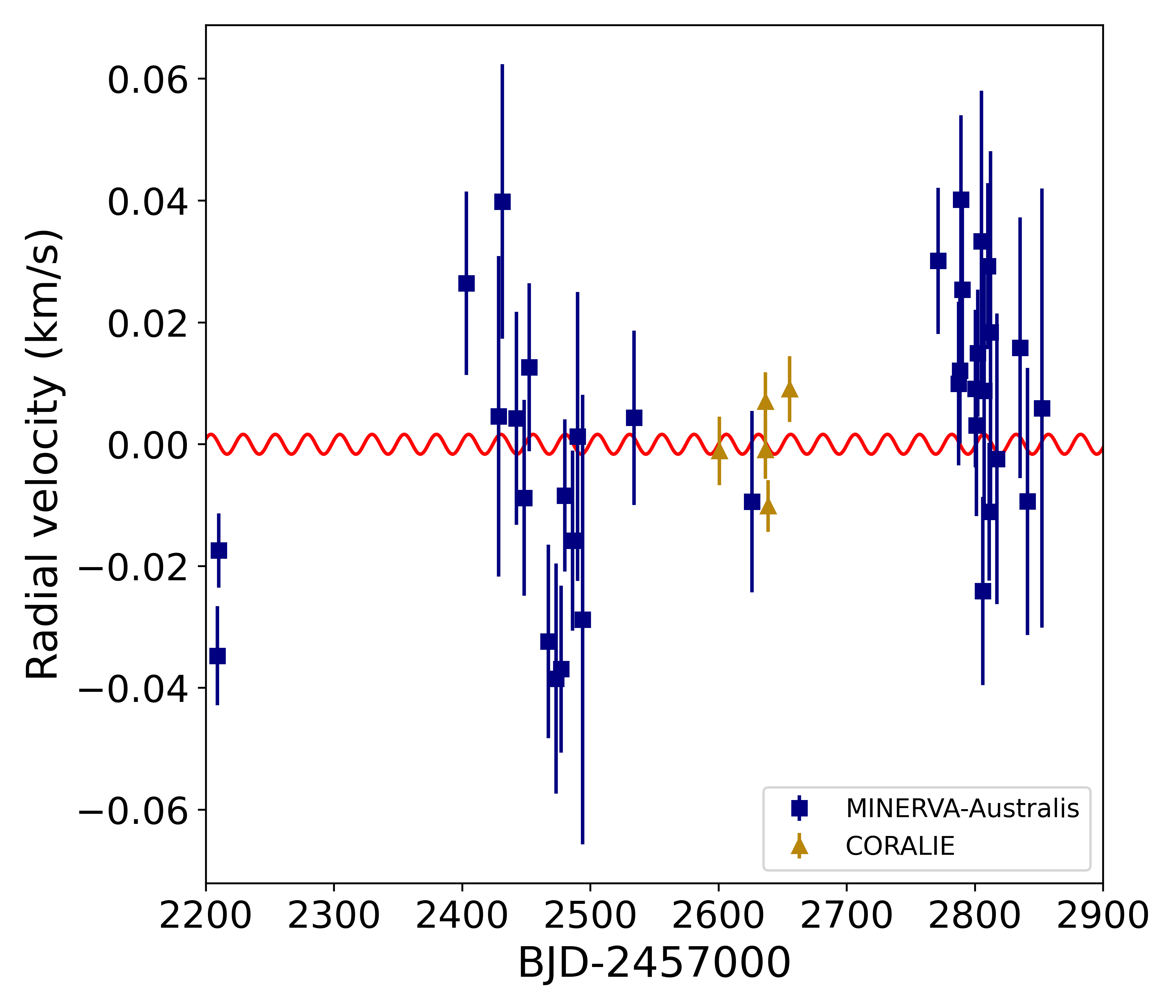}
    \caption{Radial velocity measurements of HD 21520 from \textsc{Minerva}-Australis and CORALIE and the model RV curve from the \texttt{allesfitter} fit of the \tess\, and CHEOPS photometry and ESPRESSO radial velocities.}
    \label{fig:ma_cor}
\end{figure}

We observed TOI-4320 with the high resolution CORALIE spectrograph that is installed at the Swiss 1.2-m Leonhard Euler Telescope at ESO’s La Silla Observatory \citep{2001Msngr.105....1Q}. CORALIE has a resolving power of R $\sim$ 60, 000 and is fed by a 2 arcsec fiber \citep{2010A&A...511A..45S}. A total of 5 RVs were obtained from 2022 January 21 to 2022 March 17 using exposure times of 1200 s which translated in spectra with a signal-to-noise ratio per resolution element (S/N) around 50 at 550 nm. We derived the RV of each epoch by cross-correlating the spectrum with a binary G2 mask \citep{1996A&AS..119..373B, 2002Msngr.110....9P}. These observations allow us to exclude any kind of binaries and to exclude fast rotating star. The CORALIE RVs are shown in Figure \ref{fig:ma_cor} alongside the \textsc{Minerva}-Australis RVs.

\subsection{ESPRESSO Radial Velocities\label{subsec:espresso}}

\begin{figure}[!h]
    \centering
    \includegraphics[width=\columnwidth]{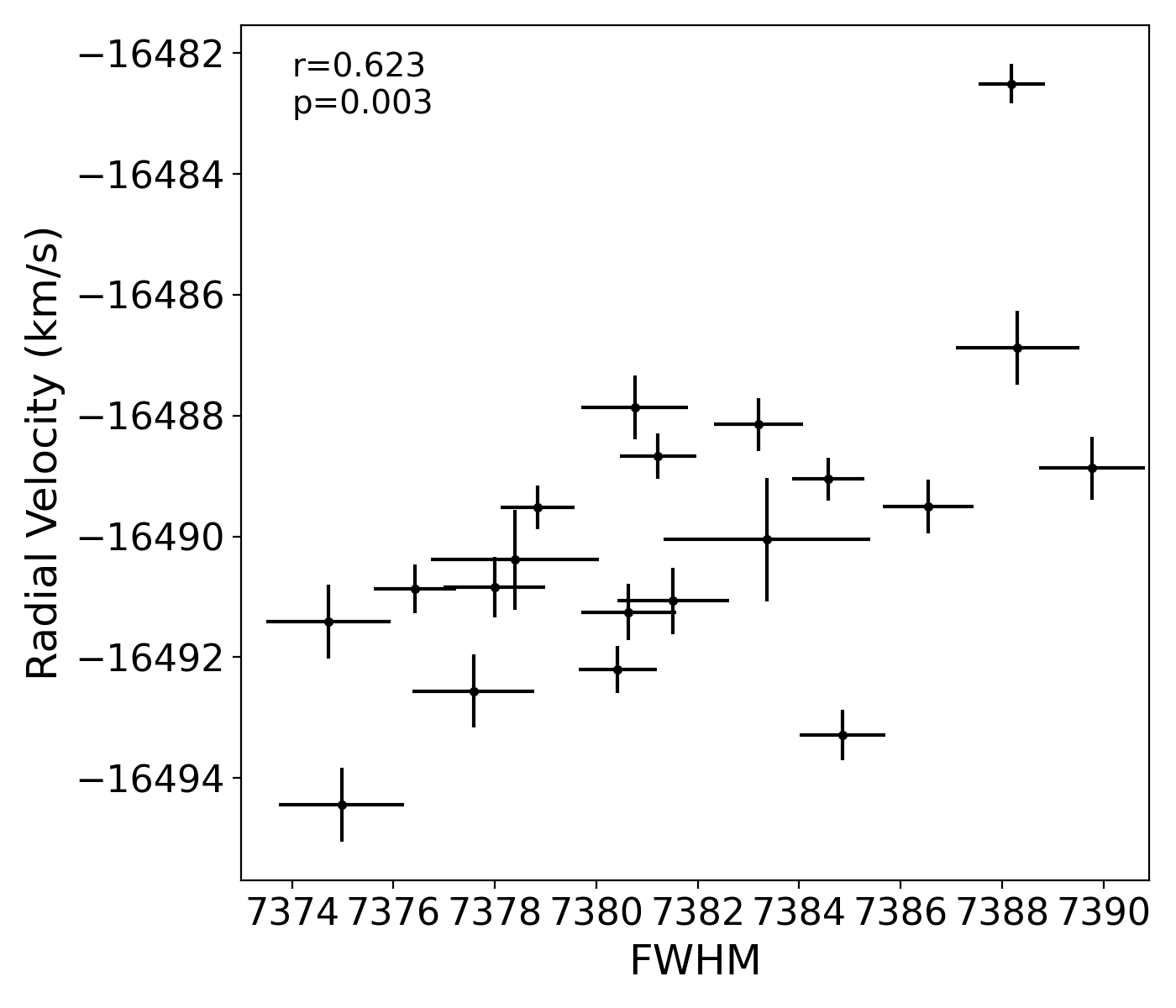}
    \caption{ESPRESSO radial velocities vs CCF$_\textrm{FWHM}$ showing a significant positive correlation, with a Pearson correlation coefficient of 0.623 (p=0.003).}
    \label{fig:fwhm_rv}
\end{figure}

\begin{figure*}[!h]
    \centering
    \includegraphics[width=\textwidth]{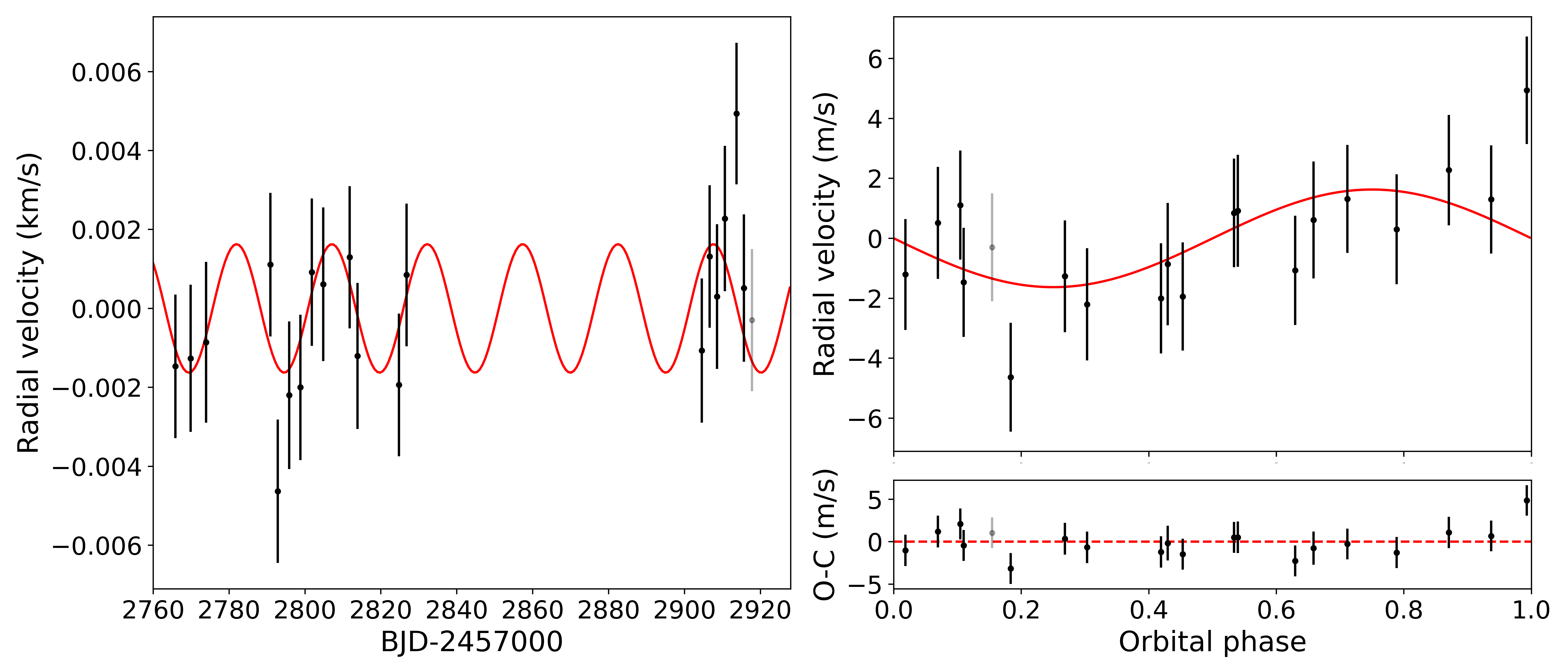}
    \caption{{\it Left:} ESPRESSO radial velocity measurements of HD 21520 and model RV curve assuming a circular orbit and best-fit mass from the \texttt{allesfitter} fit. Jitter has been added in quadrature to the measurement uncertainties. {\it Top right:} Phase-folded, offset-subtracted RVs and best-fit RV curve. {\it Bottom right:} Phase-folded residuals. The grey point in all plots is the contaminated point excluded from our analysis (see Section \ref{subsec:espresso}).}
    \label{fig:rvfit}
\end{figure*}

We acquired 21 high-resolution spectroscopic observations of TOI-4320 using ESPRESSO \citep{2021A&A...645A..96P} on the 8.2 m Very Large Telescope (VLT) located in Paranal, Chile. The observations were carried out between 2022 July 05 and 2022 December 06 as part of the observing programs 105.20P7.001, 109.23DX.001 and 110.2481.001 (PI: Bouchy), dedicated to the characterization of warm mini-Neptune transiting exoplanets. The exposure time was fixed to 600 s, a median resolving power of 140,000 using 2 $\times$ 1 binning, and a wavelength range of 380-788 nm. The RVs and activity indicators were extracted using version 3.0.0. of the ESPRESSO pipeline, and we computed the RVs by cross-correlating the Echelle spectra with a G2 numerical mask. The last data point (2022 December 06) was identified as unreliable due to strong moon light contamination with a Barycentric Earth Radial velocity very close to the stellar systemic velocity. The average uncertainty of the RV data is 0.52 m/s and the RMS is 2.90 m/s. The RV measurements are shown in Figure \ref{fig:rvfit} alongside the best-fit RV curve. We report the ESPRESSO RV measurements and their uncertainties, along with the full width at half maximum (FWHM), bisector, contrast, S-index and H$\alpha$-index in Table \ref{tab:espresso_vals}. ESPRESSO radial velocities show a significant correlation with the CCF$_\textrm{FWHM}$ due to stellar activity (see Figure \ref{fig:fwhm_rv}), with a Pearson correlation coefficient of 0.623 (p=0.003).

\subsection{WASP Observations\label{subsec:wasp}}
\begin{figure}
\includegraphics[width=\columnwidth]{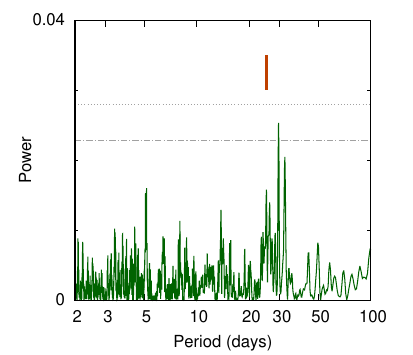}
  \caption{Lomb-Scargle Periodogram \citep{Lomb76, Scargle82} of the WASP data for HD 21520 from 2006 to 2014 combined (top). There is no significant periodicity. The horizontal lines are at the estimated 10\%\ and 1\%\ false-alarm levels. The peaks near 30 d are the residual effects of moonlight propagating through the pipeline. The red tick marks the orbital period of the planet.}
\label{fig:wasp}
\end{figure}

The WASP-South transit-search observed the field of HD 21520 from 2006 to 2011, during which time it was equipped with Canon 200-mm, f/1.8 lenses backed by 2048x2048 CCDs, observing with a 400--700 nm passband \citep{2006PASP..118.1407P}, and then continued to observe the field in 2013 and 2014 when equipped with 85-mm, f/1.8 lenses using an SDSS-$r$ filter. Observations spanned typically 150 days in each year, and in total 88\,000 photometric datapoints were recorded. HD 21520 is by far the brightest star in the extraction aperture.  We searched the accumulated dataset for any rotational modulation, both year-by-year and combining years of data, using the methods from \citet{2011PASP..123..547M}. Figure \ref{fig:wasp} shows a Generalized Lomb-Scargle Periodogram of the WASP data for HD 21520. We find no significant and persistent periodicity in the range 1 d to 100 d, with a 95\%-confidence upper limit of 0.7 mmag. 

\section{Data Analysis} \label{sec:analysis}

\subsection{Host Star Parameters\label{subsec:starparams}}

The spectroscopic stellar parameters ($T_{\mathrm{eff}}$, $\log g$, microturbulence, [Fe/H]) were derived using the ARES+MOOG methodology. This is described in detail in \citet[][]{Sousa-21, Sousa-14, Santos-13}. We used the ARES code\footnote{The last version, ARES v2, can be downloaded at https://github.com/sousasag/ARES} \citep{Sousa-07, Sousa-15} to consistently measure equivalent widths (EW) for the list of iron lines presented in \citet[][]{Sousa-08}. This spectral analysis was performed on a combined ESPRESSO spectrum for HD 21520. To converge on the best set of spectroscopic parameters we used a minimization process to find the ionization and excitation equilibrium. The process makes use of a grid of Kurucz model atmospheres \citep{Kurucz-93} and the latest version of the radiative transfer code MOOG \citep{Sneden-73}. We also derived a more accurate trigonometric surface gravity using recent Gaia data following the same procedure as described in \citet[][]{Sousa-21} which provided a consistent value when compared with the spectroscopic surface gravity (4.42 $\pm$ 0.10 dex).

\begin{figure}[!ht]
\includegraphics[width=\linewidth,trim=15 10 15 20,clip]{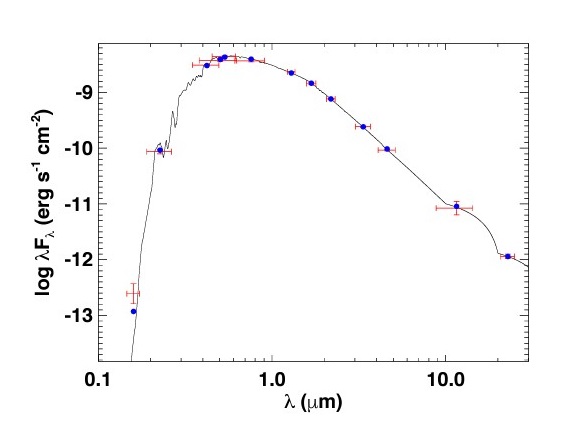}
\caption{Spectral energy distribution of HD 21520. Red symbols represent the observed photometric measurements, where the horizontal bars represent the effective width of the passband. Blue symbols are the model fluxes from the best-fit Kurucz atmosphere model (black).  \label{fig:sed}}
\label{fig:host_emission}
\end{figure}

As an independent determination of the basic stellar parameters, we performed an analysis of the broadband spectral energy distribution (SED) of the star together with the Gaia DR3 parallax \citep[with no systematic offset applied; see, e.g.,][]{StassunTorres:2021}, in order to determine an empirical measurement of the stellar radius, following the procedures described in \citet{Stassun:2016,Stassun:2017,Stassun:2018}. We pulled the $B_T V_T$ magnitudes from {\it Tycho-2}, the $JHK_S$ magnitudes from {\it 2MASS}, the W1--W4 magnitudes from {\it WISE}, the $G_{\rm BP} G_{\rm RP}$ magnitudes from Gaia, and the FUV and NUV magnitudes from {\it GALEX} (see Table \ref{tab:system_params}). Together, the available photometry spans the full stellar SED over the wavelength range 0.2--22~$\mu$m (see Figure~\ref{fig:sed}).

We performed a fit using Kurucz stellar atmosphere models, with the effective temperature ($T_{\rm eff}$), surface gravity ($\log g$), and metallicity ([Fe/H]) adopted from the spectroscopic analysis. The remaining free parameter is the extinction $A_V$, which we limited to the maximum line-of-sight value from the Galactic dust maps of \citet{Schlegel:1998}. The resulting fit (Figure~\ref{fig:sed}) has $A_V = 0.02 \pm 0.02$ with a reduced $\chi^2$ of 1.3, excluding the {\it GALEX} FUV flux which indicates a moderate level of activity (see below). Integrating the (unreddened) model SED gives the bolometric flux at Earth, $F_{\rm bol} = 5.919 \pm 0.069 \times 10^{-9}$ erg~s$^{-1}$~cm$^{-2}$. Taking the $F_{\rm bol}$ and the Gaia parallax directly gives the bolometric luminosity, $L_{\rm bol} = 1.164 \pm 0.014$~L$_\odot$, which with $T_{\rm eff}$ then gives the stellar radius, $R_\star = 1.044 \pm 0.024$~R$_\odot$. In addition, we can estimate the stellar mass from the empirical relations of \citet{Torres:2010}, giving $M_\star = 1.09 \pm 0.07$~M$_\odot$. 

% Finally, we can use the star's FUV excess (Fig.~\ref{fig:sed}) to estimate an age via empirical activity-age relations. The observed FUV excess implies a chromospheric activity of $\log R'_{\rm HK} = -4.84 \pm 0.09$ via the empirical relations of \citet{Findeisen:2011}. This estimated activity implies an age of $\tau_\star = 3.8 \pm 0.7$~Gyr via the empirical relations of \citet{Mamajek:2008}. Those relations also predict a rotation period of $18.6 \pm 3.0$~d, which is consistent with that inferred from $v\sin i$ and $R_\star$, giving $P_{\rm rot} / \sin i = 21.1 \pm 2.2$~d. 

% Edited regarding the age
Finally, we can use the star's FUV excess (Fig.~\ref{fig:sed}) to estimate an age via empirical activity-age relations. The observed FUV excess implies a chromospheric activity of $\log R'_{\rm HK} = -4.84 \pm 0.09$ via the empirical relations of \citet{Findeisen:2011}. This is consistent with the value derived from the ESPRESSO spectra of $-4.93 \pm 0.09$. This estimated activity implies an age of $\tau_\star = 5.3 \pm 0.7$~Gyr via the empirical relations of \citet{Mamajek:2008}. We add 0.1 dex systematic uncertainty in quadrature to the measurement uncertainty of $\log R'_{\rm HK}$, based on uncertainties in the absolute age scale of the clusters that calibrate the empirical relations, thus resulting in $\tau_\star = 5.3 \pm 1.1$~Gyr. Those relations also predict a rotation period of $21.4 \pm 0.9$~d. This is consistent with that inferred from $v\sin i$ and $R_\star$, namely $P_{\rm rot} / \sin i = 21.1 \pm 2.2$~d.

\begin{deluxetable}{lcc}
\tablewidth{0pc}
\tabletypesize{\scriptsize}
\tablecaption{
   System Information
    \label{tab:system_params}
}
\tablehead{
    \multicolumn{1}{c}{Parameter} &
    \multicolumn{1}{c}{Value}    &
    \multicolumn{1}{c}{Source} 
    }
\startdata
HD & 21520 & - \\
TIC & 404518509 &  TIC V8$^a$ \\
TOI & 4320 & TIC V8$^a$ \\
R.A. &  03:26:33.68 &  ExoFOP\\
Dec. & -43:36:49.19 & ExoFOP \\
$\mu_{ra}$ (mas yr$^{-1}$)  & $-2.359 \pm 0.037$ &  ExoFOP\\
$\mu_{dec}$ (mas yr$^{-1}$) & $-22.812 \pm 0.055$ & ExoFOP\\
Parallax (mas) & $12.596 \pm 0.025$ & GAIA DR3$^b$  \\
%Epoch & 1431.2679252705 $_{-0.0025105776}^{+0.0031558358}$ & This work \\
FUV (mag)    & $21.13 \pm 0.26$ & GALEX \\
NUV (mag)    & $14.44 \pm 0.10$ & GALEX \\
$B_T$ (mag)    & $9.918 \pm 0.024$ & Tycho-2 \\
$V_T$ (mag)    & $9.219 \pm 0.017$ & Tycho-2 \\
$Gaia$ (mag) &  $9.026 \pm 0.003$ & Gaia DR3 \\
$B_P$ (mag) &  $9.333 \pm 0.003$ &  Gaia DR3 \\
$R_P$ (mag) & $8.550 \pm 0.004$ &  Gaia DR3 \\
\tess\, (mag) & $8.61 \pm 0.01$ & TIC V8\\
$J$ (mag)    & $8.048 \pm 0.030$ & 2MASS$^c$  \\
$H$ (mag)    & 	$7.764 \pm 0.042$ & 2MASS \\
$K_S$ (mag)  & $7.707 \pm 0.016$ & 2MASS  \\
$T_{\rm eff}$ (K)      & $5871 \pm 62 $ & This work \\ 
$[$Fe/H$]$         & $0.049 \pm 0.042$  & This work \\
$\log g$          & $4.42 \pm 0.03$ & This work \\
$v\sin i$ (km/s) & $2.5 \pm 0.25$ & This work \\
$M_{\star}$ ($M_{\odot}$)  & $1.09 \pm 0.07$ & This work \\
$R_{\star}$ ($R_{\odot}$)   & $1.044 \pm 0.024$ & This work \\ 
$\rho_{\star}$ (g cm$^{-3}$) & $1.35 \pm 0.13$ & This work \\
$L_{\star}$ ($L_{\odot}$)   & $1.164 \pm 0.014$ & This work \\ 
$R'_{\rm HK}$ & $-4.93 \pm 0.06$ & This work \\
Age (Gyr)       &  $5.3 \pm 1.1$ & This work \\ 
\enddata
% \tablecomments{(a) \cite{2018AJ....156..102S}. (b) \cite{2018A&A...616A...1G}. (c) \cite{2003yCat.2246....0C}.}
\tablecomments{(a) \cite{2018AJ....156..102S}. (b) \cite{2023AA...674A...1G}. (c) \cite{2003yCat.2246....0C}}
\label{system_params}
\end{deluxetable}

\subsection{False Positive Scenarios and Statistical Validation\label{subsec:validation}}
To rule out possible false positive scenarios we can use \tess\ photometry in addition to the contrast curves from follow-up data. We use the \tess\ photometry with SOAR's high-resolution speckle imaging (Figure \ref{fig:soar}), and \tess\ photometry with Gemini-South high resolution imaging separately (Figure \ref{fig:gemini}). The python package \texttt{triceratops} \citep{2020ascl.soft02004G, 2021AJ....161...24G} uses lightcurves of each pixel in \tess\ target pixel files for each observed sector to calculate the probabilities of various false positive scenarios and nearby false positive scenarios. The False Positive Probability (FPP) is the probability that the observed transit is caused by something other than a transiting planet around the target star: a planet or a star transiting an unresolved bound companion or background star, or a star transiting the target star (i.e. an eclipsing binary). The Nearby False Positive Probability (NFPP) is the probability that the observed transit comes from a nearby star rather than the target star, including a nearby transiting planet or nearby eclipsing binary. The FPPs and NFPPs are constrained further with the addition of follow-up contrast curves, which we obtained from SOAR and Gemini-South. 

\begin{figure}
\includegraphics[width=0.50\textwidth]{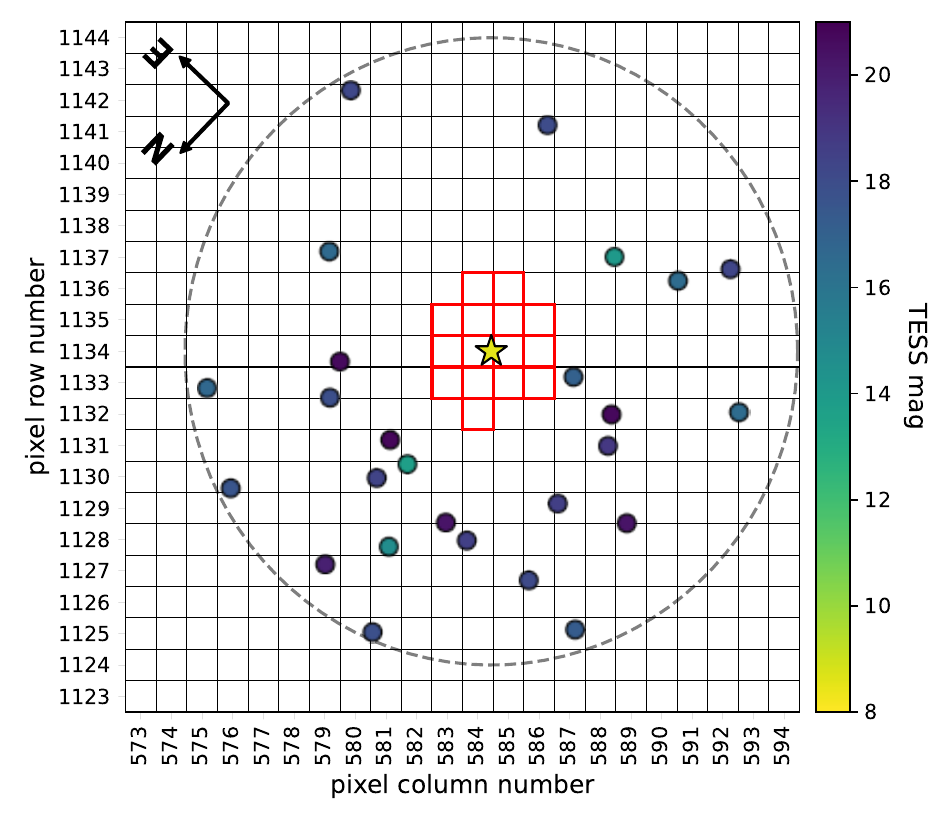}
  \caption{The \tess\ target pixel file of HD 21520 in sector 3. The pixels outlined in red are the pixels included in the aperture, and there are no nearby stars included within it. The filled colored circles are known stars from the Gaia DR2 catalog \citep{Brown18}.}
\label{fig:triceratops}
\end{figure}

In Figure \ref{fig:triceratops}, there are no other target stars within the aperture in sector 3, as with all the other sectors, which will contribute to constraining the NFPP. Due to the variation in individual calculations, we run the FPP and NFPP calculations 20 times and take the mean of the 20 calculations. With the addition of SOAR follow-up, we calculate a FPP of 0.00103 \(\pm\) (7.3537 $\times 10^{-5}$) and NFPP of 0. To be considered statistically validated, FPP \(<\) 0.015 and the NFPP \(<10^{-3}\). HD 21520 b is below the statistical validation limit within uncertainties for both the FPP and NFPP using SOAR follow-up.

Rerunning \texttt{triceratops} separately with the same \tess\ photometry data but using the 832 nm contrast curve obtained from Gemini-South high resolution imaging gives a mean FPP after 20 runs of $3.9450\times 10^{-5}$ \(\pm\) (1.251 $\times10^{-5}$) and NFPP of 0. HD 21520 b has FPPs and NFPPs well below the threshold needed for statistical validation, so we consider HD 21520 b to be statistically validated as a true planet.

In addition, we use the ESPRESSO radial velocities to obtain a 3-$\sigma$ upper mass limit for HD 21520 b of 16.8 $M_{\oplus}$, far below the stellar and brown dwarf limits.

\subsection{Fitting the Planetary Parameters of HD 21520 b}\label{subsec:planetary_params}

In order to derive the orbital and planetary parameters, we performed a joint fit of the \tess\, and \textit{CHEOPS} photometry and ESPRESSO radial velocities using the publicly available \texttt{allesfitter} package \citep{2021ApJS..254...13G}, assuming a circular orbit. We omit the contaminated ESPRESSO data point described in Section \ref{subsec:espresso}. We did not include \textsc{Minerva}-Australis radial velocities in the fit because of their increased scatter. Additionally, we do not include CORALIE radial velocities due to the small sample size. We used a nested sampling algorithm consisting of 500 walkers to explore the parameter space and determine the best-fit values for the following parameters:

\begin{itemize}
    \itemsep0pt 
    \item radius ratio, $R_p/R_{\star}$, where p denotes the individual planet, with uniform prior from 0 to 1,
    \item sum of radii divided by the orbital semi-major axis, $(R_{\star} + R_p) / a$, with uniform prior from 0 to 1,
    \item cosine of the orbital inclination, $ \cos{i}$, with uniform prior from 0 to 1, 
    \item orbital period, $P$, with uniform prior from 25.1 to 25.2,
    \item transit epoch, $T_{0}$, with uniform prior from BJD 2459159.9 to 2459160.1,
    \item radial velocity semi-amplitude, $K$, with uniform prior from 0 to 100 m/s
    \item the hyperparameters $\sigma_{GP}$ and $\rho_{GP}$ and offset$_{GP}$ for a Mat\'{e}rn 3/2 kernel used to model the red noise for the 2--minute \tess\, data
    \item white noise scaling terms for the 2-minute \tess\, data, $\sigma_{TESS}$.
    \item the hyperparameters $\sigma_{GP}$ and $\rho_{GP}$ and offset$_{GP}$ for a Mat\'{e}rn 3/2 kernel used to model the red noise for the \textit{CHEOPS} data
    \item white noise scaling terms for the 2-minute \textit{CHEOPS} data, $\sigma_{CHEOPS}$.
    \item jitter term, $\sigma$, for the ESPRESSO RVs
    \item offset term, $\gamma$, for the ESPRESSO RVs

\end{itemize}

\begin{figure}
\includegraphics[width=0.50\textwidth]{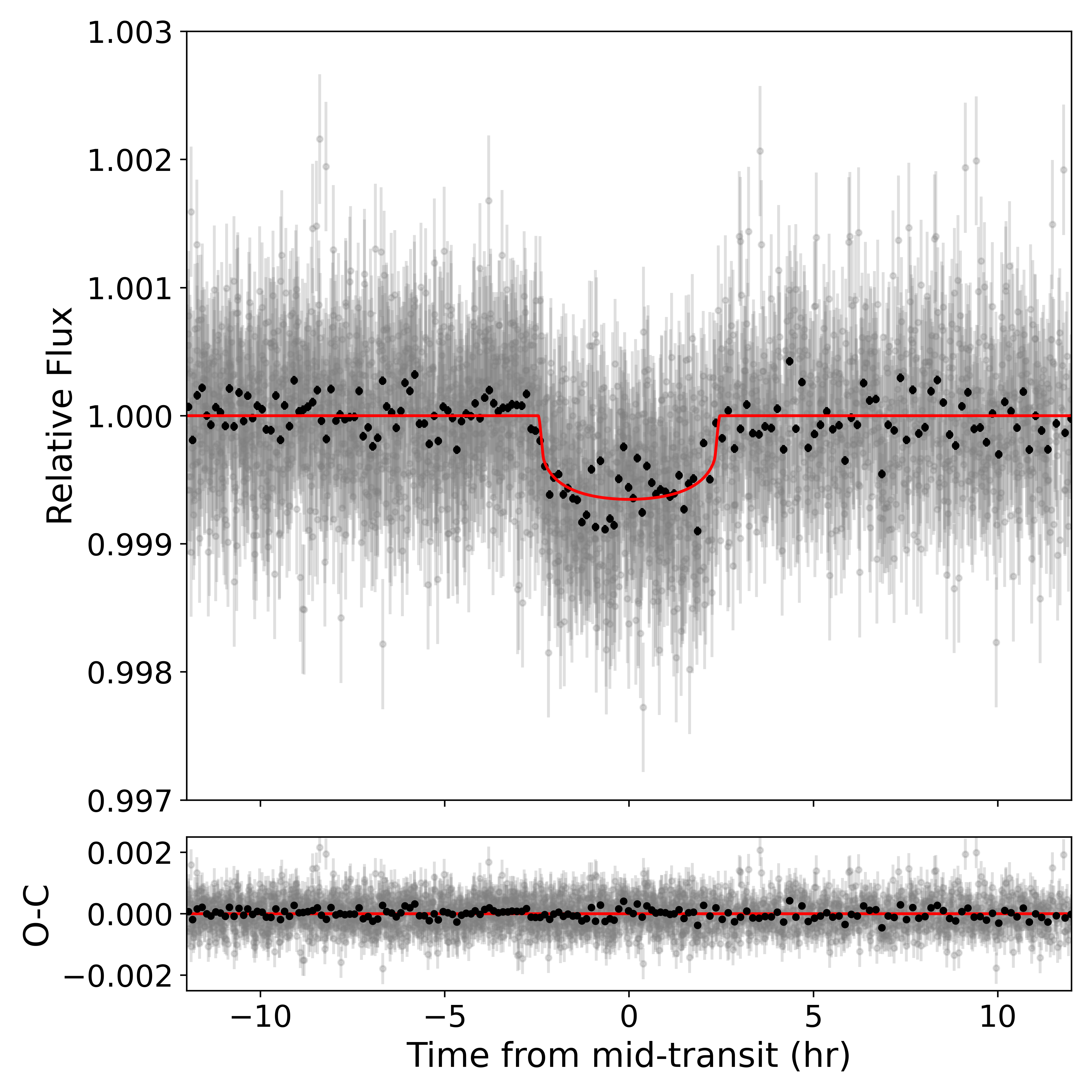}
    \caption{Phase folded \tess\, transit data with modeled \texttt{allesfitter} fit (grey), and residuals plot on the bottom. Also plotted are the phase folded points binned to 30 minutes (black).}  
\label{fig:tess_phased}
\end{figure}

The values of $q_1$ and $q_2$ for both \tess\, and \textit{CHEOPS} were obtained by matching the spectroscopic parameters of the primary star to the closest values of the coefficients $u_1$ and $u_2$ of the quadratic limb darkening law listed in \citet{Claret17}, and transforming them to the corresponding values of $q_1$ and $q_2$. These values were fixed for the fit and are listed in Table \ref{tab:planetary_params}. Due to the aforementioned correlation between the ESPRESSO RVs and CCF$_\textrm{FWHM}$ (see Section \ref{subsec:espresso}), we detrend the RVs against the CCF$_\textrm{FWHM}$ prior to fitting. The values and uncertainties of the fitted and derived parameters listed in Table \ref{tab:planetary_params} are defined as the median values and 68\% confidence intervals of the posterior distributions, respectively. The best-fit phase folded transit model is shown alongside the CHEOPS and \tess\, data  in Figures \ref{fig:cheops_phot} and \ref{fig:tess_phased}, respectively. The best-fit RV model and ESPRESSO data is shown in Figure \ref{fig:rvfit}. The corner plots for the modeled and derived parameters are shown in Figure \ref{fig:corner_fitted} and Figure \ref{fig:corner_derived} of the Appendix.

Regarding the effective temperature reported using \texttt{allesfitter}, this temperature is reported as a lower limit. It is more likely that HD 21520 b is tidally locked due to its orbital period \citep{Peale77}, so if we assume that HD 21520 b is tidally locked and has no heat re-circulation, we get an upper limit equilibrium temperature of about 832 K.

\begin{deluxetable}{lcc}
\tablewidth{0pc}
\tabletypesize{\scriptsize}
\tablecaption{
    Planetary Parameters
    \label{tab:planetary_params}
}
\tablehead{
    \multicolumn{1}{c}{Parameter} &
    \multicolumn{1}{c}{Value}    &
    \multicolumn{1}{c}{Error} 
    }
\startdata
\textit{Fixed Parameters} &  & \\
$q_{1,\mathrm{TESS}}$ & $0.32$ &  - \\ 
$q_{2,\mathrm{TESS}}$ & $0.30$ &  - \\
$q_{1,\mathrm{CHEOPS}}$ & $0.47$ &  - \\ 
$q_{2,\mathrm{CHEOPS}}$ & $0.36$ &  - \\
$\sqrt{e}\cos{\omega}$ & $0.0$ & - \\ 
$\sqrt{e}\sin{\omega}$ & $0.0$ &  - \\
 & & \\
\textit{Modeled Parameters} & & \\
$(R_\star + R_p) / a$ & $0.02878$ &$_{-0.00090}^{+0.0010}$\\ 
$R_p / R_\star$ & 0.02369 & $\pm 0.00056$\\ 
$\cos{i}$ & $0.0134$ & $\pm 0.0022$ \\ 
$T_{0}$ JD & 2459160.0157 & $\pm 0.0015$ \\
$P(\mathrm{d})$ & $25.1292$ &  $_{-0.000073}^{+0.000089}$ \\ 
$K (m/s)$ & 1.64 & $\pm 0.63$ \\
$\log{\sigma_\mathrm{TESS}}$ & $-7.591$ & $\pm0.014$ \\ 
offset$_\mathrm{TESS}$ & $0.000002$ & $\pm 0.000028$ \\ 
$\ln{\sigma_\mathrm{TESS}}$ & $-9.84$ & $_{-0.50}^{+0.56}$  \\ 
$\ln{\rho_\mathrm{TESS}}$ & $-0.4$ & $_{-1.1}^{+4.9}$  \\ 
$\log{\sigma_\mathrm{CHEOPS}}$ & $-8.019$ & $\pm0.025$  \\ 
offset$_\mathrm{CHEOPS}$ & $0.000280$ & $_{-0.000037}^{+0.000041}$  \\ 
$\ln{\sigma_\mathrm{CHEOPS}}$ & $-8.165$ & $\pm 0.091$  \\ 
$\ln{\rho_\mathrm{CHEOPS}}$ & $-5.83$ & $_{-0.22}^{+0.24}$  \\ 
$\ln{\sigma_\mathrm{jitter, ESPRESSO}}$ [km/s] & $-6.34$ & $_{-0.17}^{+0.19}$ \\ 
$\gamma_{ESPRESSO}$ (m/s) & $0.14$ & $\pm 0.41$ \\ 
 & & \\
\textit{Derived Properties} & & \\
$R_\star/a$ & $0.02811$ &  $_{-0.00088}^{+0.00100}$ \\ 
$a/R_\star$ & $35.6$ &  $_{-1.2}^{+1.1}$\\ 
$R_\mathrm{p}/a$ & $0.000666$ &  $_{-0.000029}^{+0.000032}$ \\ 
$R_\mathrm{p}$ ($\mathrm{R_{\oplus}}$) & $2.697$ &  $\pm0.090$ \\ 
%$R_\mathrm{p}$ ($\mathrm{R_{jup}}$) & $0.240$ &  $\pm0.011$ \\ 
$M_\mathrm{p}$ ($\mathrm{M_{\oplus}}$) & $7.9^{a}$ & $_{-3.0}^{+3.2}$ \\ %23.7 & $_{-16.3}^{+24.0}$ \\
%$M_\mathrm{p}$ ($\mathrm{M_{jup}}$) & $<0.35^{a}$ & - \\
%$a$ ($\mathrm{R_{\odot}}$) & $37.4$ &  $\pm2.3$\\ 
$a$ (AU) & $0.1726$ &  $\pm0.0071$\\ 
$i$ (deg) & $89.23$ & $\pm0.12$ \\ 
e & 0 (fixed) & - \\
$b$ & $0.478$ &  $_{-0.066}^{+0.058}$ \\ 
$T_\mathrm{tot}$ (h) & $4.885$ & $_{-0.075}^{+0.083}$ \\ 
$T_\mathrm{full}$ (h) & $4.592$ &  $_{-0.083}^{+0.090}$ \\ 
$\rho_\mathrm{\star;b}$ (g cm$^{-3}$) & $1.35$ &  $\pm 0.13$\\ 
$T_\mathrm{eq}(K)^{b} $ & 637 & $_{-12}^{+13}$ \\ %Assuming an albedo of 0.3 and emissivity of 1.
$\delta_\mathrm{tr; undil; b; tess}$ (ppt) & $0.631$ & $_{-0.027}^{+0.029}$ \\ 
$\delta_\mathrm{tr; dil; b; tess}$ (ppt) & $0.631$ & $_{-0.027}^{+0.029}$ \\
$\delta_\mathrm{tr; undil; b; cheops}$ (ppt) & $0.654$ & $_{-0.026}^{+0.029}$ \\ 
$\delta_\mathrm{tr; dil; b; cheops}$ (ppt) & $0.654$ & $_{-0.026}^{+0.029}$\\
\enddata
\tablecomments{(a) 3-$\sigma$ upper limit of 17.7 $M_{\oplus}$. (b)Assuming an albedo of 0.3 and emissivity of 1.}
\label{planetary_params}
\end{deluxetable}

\section{Discussion} \label{sec:discussion}
\begin{figure}
\includegraphics[width=\columnwidth]{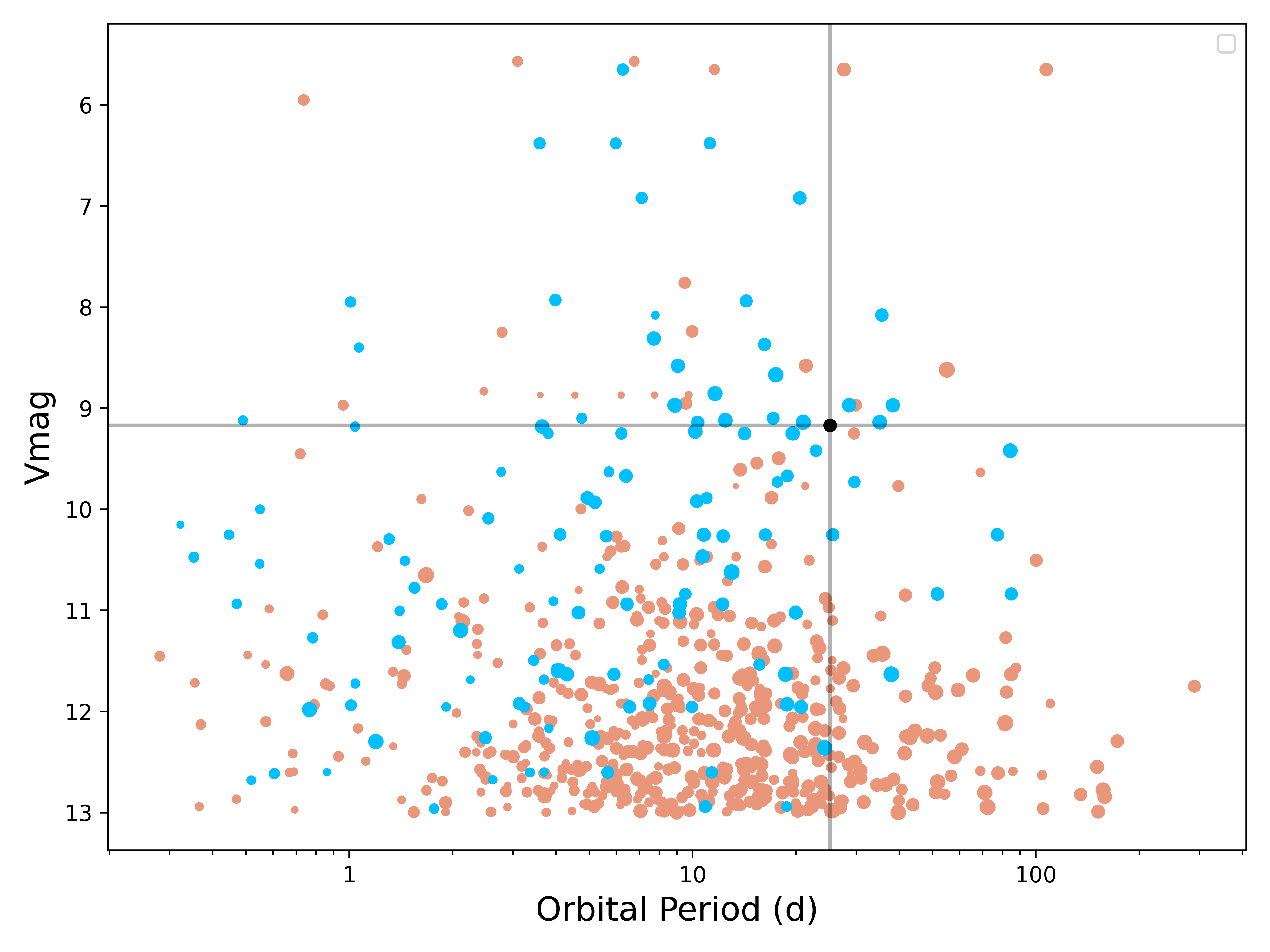}
  \caption{Orbital period versus V magnitude for sub-Neptune ($<4\, R_{\oplus}$) transiting exoplanet systems, with blue denoting a \tess\ discovery and orange being a non-\tess\ discovery. The black dot denotes the HD 21520 system. The marker sizes denote the relative radii of the planets.}
\label{fig:comp}
\end{figure}

We have determined that HD 21520 b has a period of 25.13 days and radius $2.70\pm{0.09} R_{\oplus}$, which puts it in the mid-range period sub-Neptune category. In Figure \ref{fig:comp}, we compare HD 21520 b to other transiting planets with host stars having Vmag $<$ 13 and radius less than 4$R_{\oplus}$. HD 21520 b is among a group of mid-range period planets with brighter host stars than most planets discovered with similar periods. Additionally, we obtain a 1-$\sigma$ mass measurement of $7.9^{+3.2}_{-3.0}$ $M_{\oplus}$ and 3-$\sigma$ upper mass limit of 17.7 $M_{\oplus}$.

HD 21520 b is added to the group of just 15 systems of transiting sub-Neptune planets with periods $>$ 20 days and host stars of Vmag $<$ 10 (NASA Exoplanet Archive). Interestingly, of those all but three (Kepler-409 b, \citealt{Morton16}; HD 95338 b, \citealt{Diaz20}; HD 56414 b, \citealt{Giacalone22}) are known to have multiple transiting planets. Whether HD 21520 b will remain in this small subgroup as another lone warm sub-Neptune remains to be determined. HD 21520 is characteristically similar to our Sun, with effective temperature 5858 K, radius 1.04 $R_{\odot}$, and mass of 1.09 $M_{\odot}$. Assuming the above value for HD 21520 b's mass and using the equilibrium temperature value from Table \ref{tab:planetary_params}, we find a Transmission Spectroscopy Metric (TSM; \citealt{Kempton18}) value of 39. If we instead assume the more realistic (i.e. no heat redistribution) equilibrium temperature of 832 K, we obtain a TSM value of 52. We conclude that HD 21520 b will be a promising target for atmospheric characterization with transmission spectroscopy. We note that even though transiting sub-Neptunes with equilibrium temperature comparable to that of HD 21520 b are also known to orbit with periods shorter than 20 days around lower mass (K or M) stars, we expect their properties (such as atmospheric metallicity and overall composition) to differ from those of warm sub-Neptunes around G stars, such as HD 21520 b. These differences could be driven by a number of factors, including different protoplanetary disk composition and mass \citep{Mah23,Pas16,Pas09}, as well as enhanced atmospheric erosion due to a higher level of stellar activity and flares for planets around lower mass stars \citep{John21}. Thus, future studies of HD 21520 b will enable the planet to serve as a valuable point of comparison both against planets around the Sun, as well as planets of similar size and temperature orbiting K and M dwarfs.

\begin{figure}
\includegraphics[width=\columnwidth]{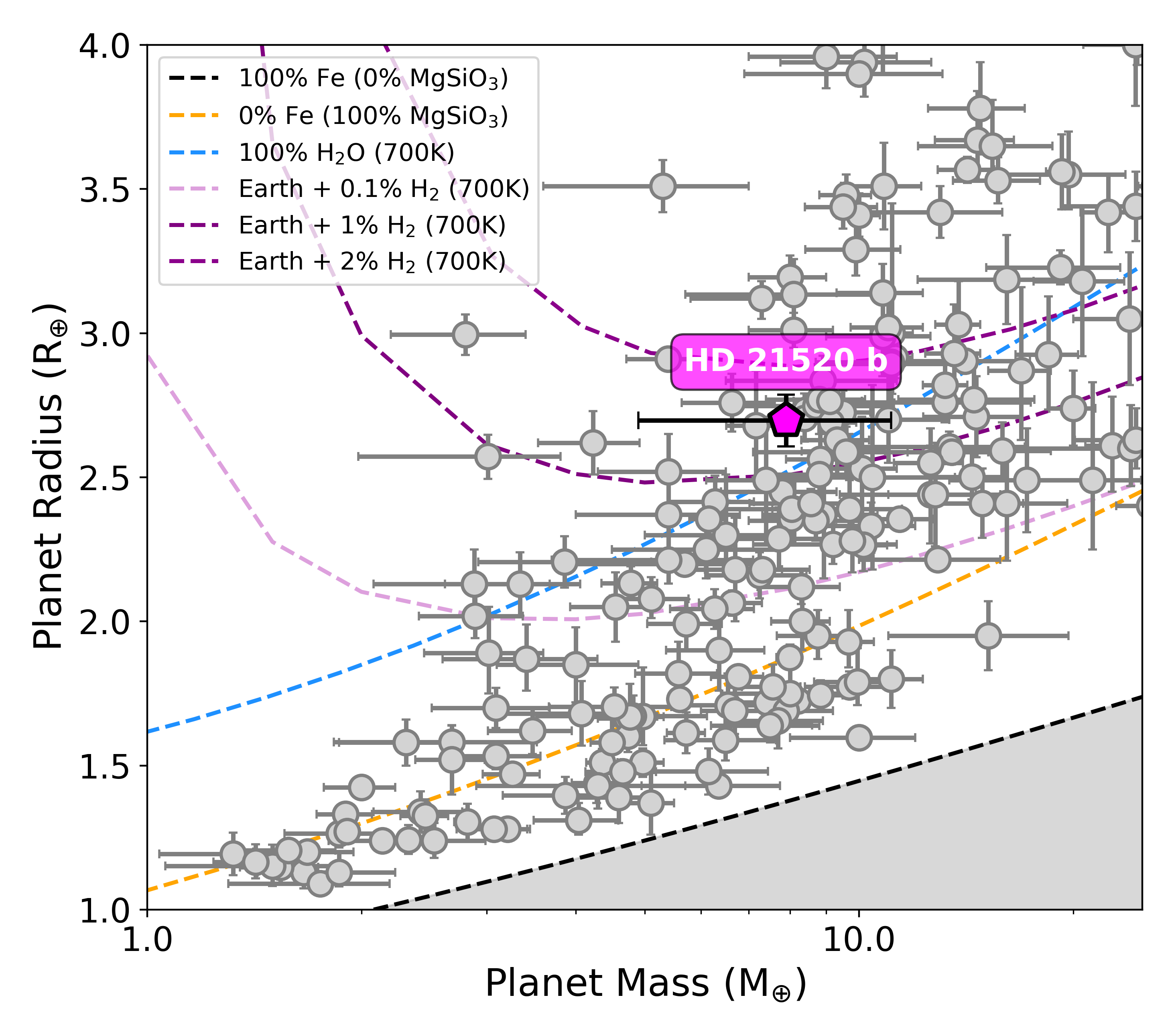}
  \caption{Mass-radius diagram made using \texttt{mr-plotter} \citep{2023arXiv230504922C} of sub-Neptunes with measured masses as well as theoretical models from \citet{2019PNAS..116.9723Z}. Only planets with 3-$\sigma$ mass and radius measurements are shown for clarity. The position of HD 21520 b on the diagram suggests its composition could be that of a rocky core surrounded by a H$_2$ envelope, although a ``water-world" composition is possible given the large uncertainty on the mass measurement.}
\label{fig:massrad}
\end{figure}

\begin{figure*}
\includegraphics[width=0.49\textwidth]{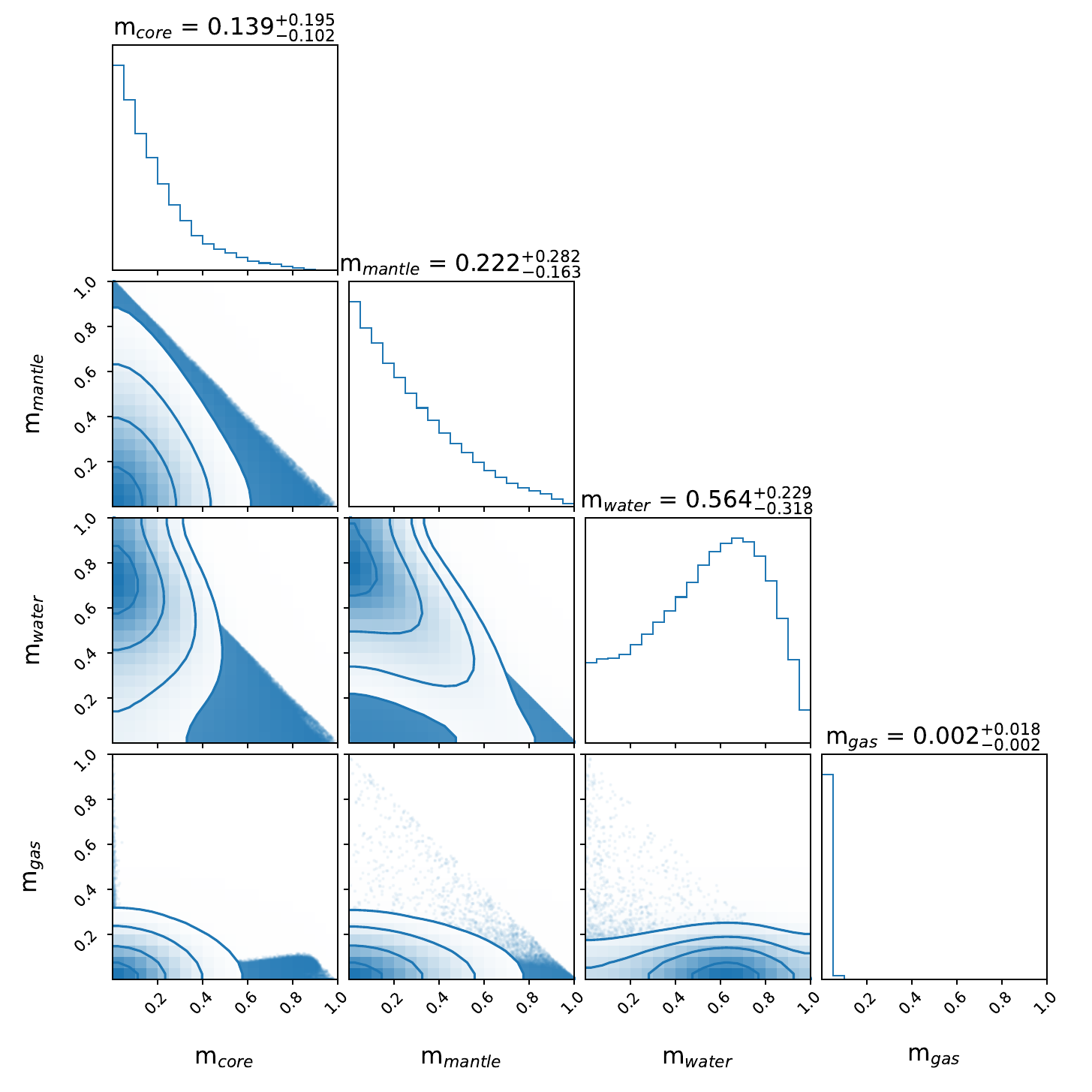}
\includegraphics[width=0.49\textwidth]{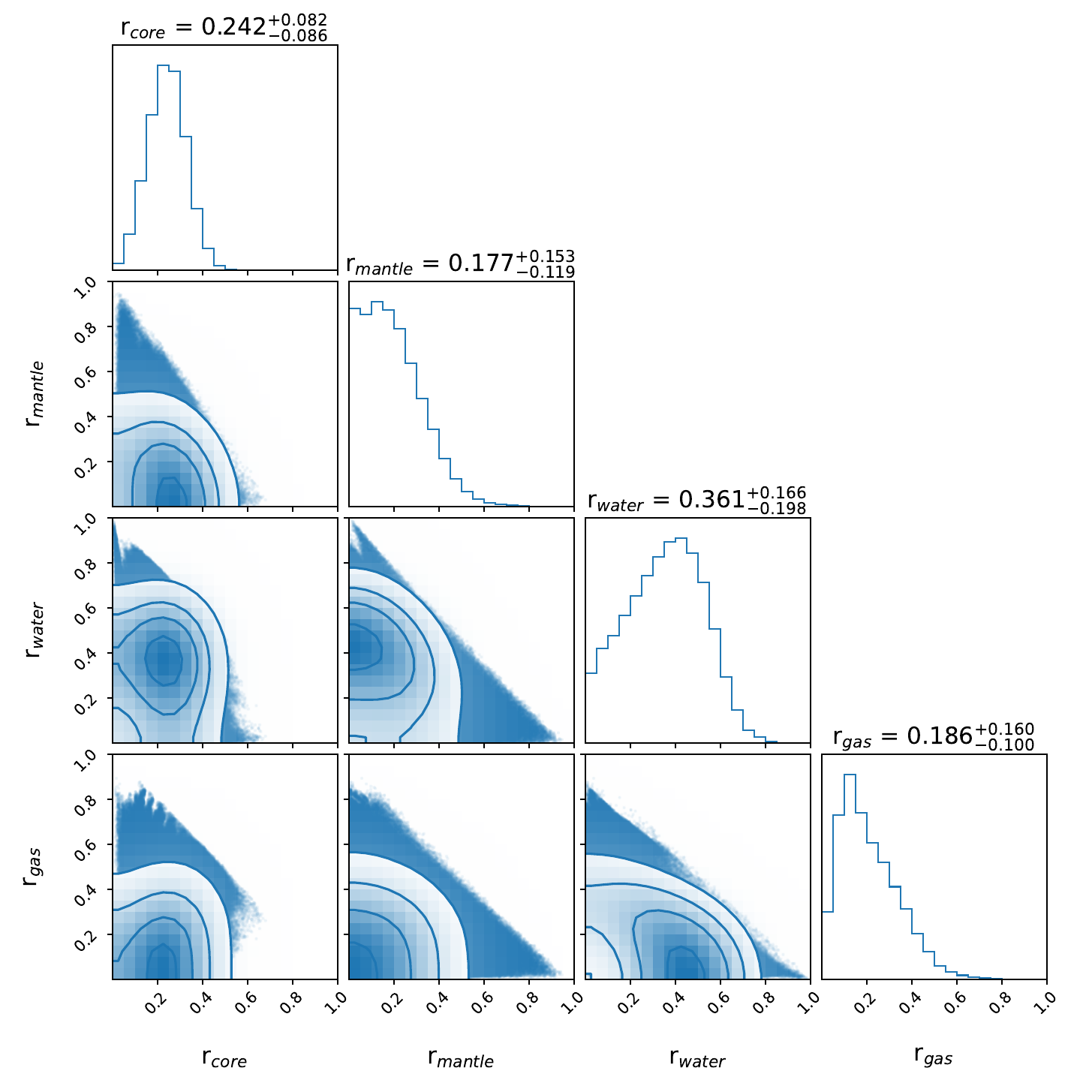}
  \caption{\textit{Left:} Mass fractions for the core, mantle, water, and gas layers calculated using ExoMDN. \textit{Right:} Radius fractions for the same layers.}
\label{fig:exomdn}
\end{figure*}

HD 21520 b's bulk density of $2.21^{+0.43}_{-0.41}$ g/cm$^3$ implies that it likely contains a significant atmosphere (Figure \ref{fig:massrad}). We use the machine-learning tool ExoMDN \citep{2023A&A...676A.106B} to model the possible interior structure of the planet, and determine the mass and radius fractions for the core, mantle, water, and gas layers. We run 1000 predictions using radius, mass, and temperature values drawn from normal distributions centered on the best-fit values, and draw 5000 samples for each prediction. ExoMDN predicts that planet contains a significant water mass and radius fraction, as well as a significant gas radius fraction (Figure \ref{fig:exomdn}). However, there are large uncertainties in these values given the low significance of our mass measurement. Thus, more RV measurements are needed to refine the mass and improve predictions about the structure of the planet, which in turn will help better inform any future atmospheric characterization.

\section{Summary} \label{sec:summary}
In this paper, we have validated the planetary nature of HD 21520 b as a sub-Neptune planet ($2.70\pm{0.09}\, R_{\oplus}$) orbiting a bright G star, with period of 25.13 days. Additionally, we are able to make a tentative mass measurement of $7.9^{+3.2}_{-3.0}\, M_{\oplus}$, with a 3-$\sigma$ upper limit of $17.7\, M_{\oplus}$. HD 21520 b is the newest addition to the group of intermediate period planets orbiting bright stars. In order to determine physical and orbital properties for this transiting planet, we used \texttt{allesfitter} to determine the physical and orbital planetary parameters. In addition, we use \texttt{triceratops} to statistically validate HD 21520 b as a true planet. Ground based follow-up observations also point to the absence of nearby eclipsing binaries and confirms that HD 21520 b is indeed a planet. We thus validate the planetary nature of HD 21520 b, and present it as a promising candidate for atmospheric characterization - particularly through transmission spectroscopy - due its size, likely low mass, and bright host star.


\facilities{\tess, CHEOPS LCOGT, SOAR, Gemini-South, \textsc{Minerva}-Australis, ESPRESSO, CORALIE, WASP}

\software{AstroImageJ \citep{Collins:2017}, \texttt{triceratops} \citep{2020ascl.soft02004G, 2021AJ....161...24G}, \texttt{allesfitter} \citep{2021ApJS..254...13G}, BLS \citep{Kovacs02}}

%\begin{acknowledgments}
\section*{Acknowledgments}

We thank the anonymous reviewer for their helpful comments, which have helped improve the paper.

DD acknowledges support from the TESS Guest Investigator program under NASA grants 80NSSC21K0108 and 80NSSC22K0185. CAC acknowledges that this research was carried out at the Jet Propulsion Laboratory, California Institute of Technology, under a contract with the National Aeronautics and Space Administration (80NM0018D0004). KAC acknowledges support from the TESS mission via subaward s3449 from MIT.

We thank the Swiss National Science Foundation (SNSF) and the Geneva University for their continuous support to our planet low-mass companion search programs. This work has been carried out within the framework of the National Centre of Competence in Research PlanetS supported by the Swiss National Science Foundation.  

Some of the observations in this paper made use of the High-Resolution Imaging instrument Zorro and were obtained under Gemini LLP Proposal Number: GN/S-2021A-LP-105. Zorro was funded by the NASA Exoplanet Exploration Program and built at the NASA Ames Research Center by Steve B. Howell, Nic Scott, Elliott P. Horch, and Emmett Quigley. Zorro was mounted on the Gemini South telescope of the international Gemini Observatory, a program of NSF’s OIR Lab, which is managed by the Association of Universities for Research in Astronomy (AURA) under a cooperative agreement with the National Science Foundation. on behalf of the Gemini partnership: the National Science Foundation (United States), National Research Council (Canada), Agencia Nacional de Investigación y Desarrollo (Chile), Ministerio de Ciencia, Tecnología e Innovación (Argentina), Ministério da Ciência, Tecnologia, Inovações e Comunicações (Brazil), and Korea Astronomy and Space Science Institute (Republic of Korea).

We acknowledge the use of public TESS data from pipelines at the TESS Science Office and at the TESS Science Processing Operations Center. 
 
Resources supporting this work were provided by the NASA High-End Computing (HEC) Program through the NASA Advanced Supercomputing (NAS) Division at Ames Research Center for the production of the SPOC data products.
               
This paper made use of data collected by the TESS mission and are publicly available from the Mikulski Archive for Space Telescopes (MAST) operated by the Space Telescope Science Institute (STScI).
 
Funding for the TESS mission is provided by NASA’s Science Mission Directorate.

This work makes use of observations from the LCOGT network. Part of the LCOGT telescope time was granted by NOIRLab through the Mid-Scale Innovations Program (MSIP). MSIP is funded by NSF.

\textsc{Minerva}-Australis is supported by Australian Research Council LIEF Grant LE160100001, Discovery Grants DP180100972 and DP220100365, Mount Cuba Astronomical Foundation, and institutional partners University of Southern Queensland, UNSW Sydney, MIT, Nanjing University, George Mason University, University of Louisville, University of California Riverside, University of Florida, and The University of Texas at Austin.

We respectfully acknowledge the traditional custodians of all lands throughout Australia, and recognise their continued cultural and spiritual connection to the land, waterways, cosmos, and community. We pay our deepest respects to all Elders, ancestors and descendants of the Giabal, Jarowair, and Kambuwal nations, upon whose lands the \textsc{Minerva}-Australis facility at Mt Kent is situated.

The contributions of ML, FB, XD, NG, BL and CL have been carried out within the framework of the NCCR PlanetS supported by the Swiss National Science Foundation under grants 51NF40\_182901 and 51NF40\_205606. ML acknowledges support of the Swiss National Science Foundation under grant number PCEFP2\_194576.
%\end{acknowledgments}

%% p



%\appendix

%\section{TBD}

\bibliography{main}{}
\bibliographystyle{aasjournal}



\appendix
\section{LCOGT Transit Non-detections}
\restartappendixnumbering
In this appendix, we show the LCOGT observations scheduled when the system was thought to have 2 planet candidates.

%\begin{appendix}
\begin{figure*}[!h]
    \centering
    \includegraphics[width=\textwidth]{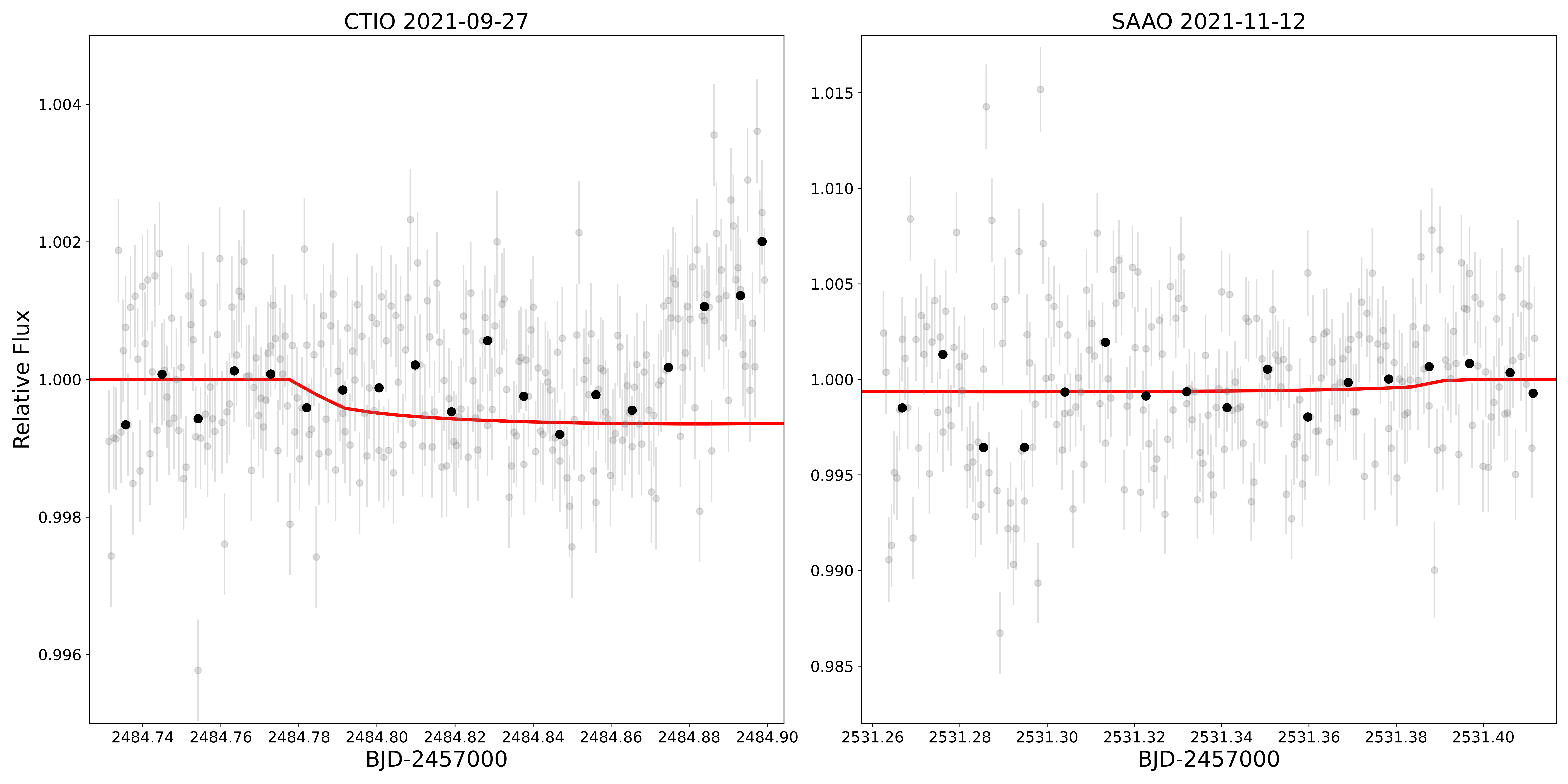}
    \caption{Unbinned (gray) and binned (black) LCO observations taken on 2021-09-27 (left) and 2021-11-12 (right) for the alleged 46.4 day candidate showing no transit. The red line denotes the best-fit transit model, but with the timing changed to that expected of the 46.4 day candidate.}
    \label{fig:lco_nontransits}
\end{figure*}

\section{ESPRESSO RVs and Activity Indices}
\restartappendixnumbering
In this appendix, we present the values of the ESPRESSO radial velocities and activity indices with uncertainties.

\begin{deluxetable}{llllllll}
\tablewidth{0pc}
\tabletypesize{\scriptsize}
\tablecaption{
    RVs and activity indices obtained from the ESPRESSO spectra
    \label{tab:espresso_vals}
}
\tablehead{
    \multicolumn{1}{l}{BJD-2400000} &
    \multicolumn{1}{l}{RV [m/s]} &
    \multicolumn{1}{l}{FWHM} & 
    \multicolumn{1}{l}{Bisector} &
    \multicolumn{1}{l}{Contrast} &
    \multicolumn{1}{l}{S$_{MW}$} &
    \multicolumn{1}{l}{H$_{\alpha}$} &
    \multicolumn{1}{l}{log(R'$_{HK}$)}
    }
\startdata
%\begin{table}[]
%\begin{tabular}{lllllllllllllll}
%BJD-2400000 & RV [m/s] & sigma_rv & FWHM & sigmafwhm & Bisector & sigbisspan & Contrast & sigcontrast & S$_{mw}$ & sigs & H$_{\alpha}$ & sigha & log(R'$_{HK}$) & sigrhk  \\
59765.8698 & -16489.507 $\pm$ 0.448 & 7386.54  $\pm$ 0.90   & -41.52 $\pm$ 0.90  & 52.9524 $\pm$ 0.0064     & 0.165449 $\pm$ 9$\times 10^{-5}$    & 0.198755     $\pm$ 3.5$\times 10^{-5}$ & -4.91052 $\pm$ 0.00050 \\
59769.8585   & -16492.557 $\pm$ 0.601 & 7377.58 $\pm$ 1.20   & -44.32 $\pm$ 1.20     & 53.0685 $\pm$ 0.0086     & 0.155419 $\pm$ 0.00015  & 0.197575 $\pm$ 5.1$\times 10^{-5}$ & -4.96936 $\pm$ 0.00094 \\
59773.9090 & -16490.053 $\pm$ 1.018 & 7383.36 $\pm$ 2.04   & -37.28 $\pm$ 2.04     & 53.0510 $\pm$ 0.0146    & 0.137386 $\pm$ 0.000321 & 0.200305 $\pm$ 0.0001  & -5.10061 $\pm$ 0.00273 \\
59790.8637 & -16488.146 $\pm$ 0.438 & 7383.19 $\pm$ 0.88    & -41.03 $\pm$ 0.87      & 52.9452 $\pm$ 0.0063     & 0.167683 $\pm$ 8.1$\times 10^{-5}$  & 0.199997 $\pm$ 3.5$\times 10^{-5}$ & -4.89844 $\pm$ 0.00043 \\
59792.8451 & -16493.284 $\pm$ 0.420 & 7384.86 $\pm$ 0.84   & -39.09 $\pm$ 0.84     & 52.9206 $\pm$ 0.0060     & 0.168326 $\pm$ 7.4$\times 10^{-5}$  & 0.199394 $\pm$ 3.3$\times 10^{-5}$ & -4.89502 $\pm$ 0.00039 \\
59795.8622   & -16494.438  $\pm$ 0.616 & 7374.98 $\pm$ 1.23  & -43.25 $\pm$ 1.23     & 53.0516 $\pm$ 0.0089     & 0.156242 $\pm$ 0.000142 & 0.197516 $\pm$ 5.5$\times 10^{-5}$ & -4.96421 $\pm$ 0.00088 \\
59798.7773 & -16488.869 $\pm$ 0.521 & 7389.77 $\pm$ 1.04   & -39.02 $\pm$ 1.04   & 52.9360 $\pm$ 0.0075     & 0.164567 $\pm$ 0.000114 & 0.198128 $\pm$ 4.3$\times 10^{-5}$ & -4.91539 $\pm$ 0.00063 \\
59801.7969 & -16491.413 $\pm$ 0.614 & 7374.72 $\pm$ 1.23   & -38.99 $\pm$ 1.23     & 53.0333 $\pm$ 0.0088     & 0.157265 $\pm$ 0.000148 & 0.198837 $\pm$ 5.3$\times 10^{-5}$ & -4.95791 $\pm$ 0.00090 \\
59804.7841 & -16490.385 $\pm$ 0.829 & 7378.40 $\pm$ 1.66   & -45.46 $\pm$ 1.66     & 53.0202 $\pm$ 0.0119      & 0.145661 $\pm$ 0.000241 & 0.196654     $\pm$ 7.6$\times 10^{-5}$ & -5.03543 $\pm$ 0.00176 \\
59811.7843 & -16488.676 $\pm$ 0.375 & 7381.22 $\pm$ 0.75  & -39.68 $\pm$ 0.75    & 52.9151 $\pm$ 0.0054     & 0.169227 $\pm$ 6$\times 10^{-5}$    & 0.200039 $\pm$ 2.9$\times 10^{-5}$ & -4.89028  $\pm$ 0.00031 \\
59813.8272    & -16491.068 $\pm$ 0.552 & 7381.52 $\pm$ 1.10    & -40.69 $\pm$ 1.10 & 52.9942 $\pm$ 0.0079     & 0.164384 $\pm$ 0.000115 & 0.199382 $\pm$ 4.9$\times 10^{-5}$ & -4.91640 $\pm$ 0.00064  \\
59824.7626   & -16492.201 $\pm$ 0.386 & 7380.42  $\pm$ 0.77   & -40.95 $\pm$ 0.77     & 52.9514 $\pm$ 0.0055     & 0.166846 $\pm$ 6.2$\times 10^{-5}$  & 0.197586 $\pm$ 3$\times 10^{-5}$   & -4.90292  $\pm$ 0.00034 \\
59826.7789 & -16490.866 $\pm$ 0.404 & 7376.42  $\pm$ 0.81   & -42.43 $\pm$ 0.81     & 52.9843 $\pm$ 0.0058     & 0.167706 $\pm$ 6.7$\times 10^{-5}$  & 0.197471 $\pm$ 3.2$\times 10^{-5}$ & -4.89831 $\pm$ 0.00036  \\
59904.5804 & -16491.252 $\pm$ 0.468 & 7380.64 $\pm$ 0.94   & -39.98 $\pm$ 0.94    & 52.9826 $\pm$ 0.0067     & 0.165643 $\pm$ 8.9$\times 10^{-5}$  & 0.199846 $\pm$ 3.8$\times 10^{-5}$ & -4.90946 $\pm$ 0.00049 \\
59906.6367 & -16489.518 $\pm$ 0.368 & 7378.84 $\pm$ 0.74   & -42.17 $\pm$ 0.74     & 52.9536 $\pm$ 0.0053     & 0.167911 $\pm$ 5.7$\times 10^{-5}$  & 0.198587 $\pm$ 2.9$\times 10^{-5}$ & -4.89722 $\pm$ 0.00030 \\
59908.5732 & -16490.842 $\pm$ 0.498 & 7377.99 $\pm$ 1.00   & -43.21 $\pm$ 1.00      & 52.9792 $\pm$ 0.0071     & 0.16244 $\pm$ 0.000103 & 0.196937 $\pm$ 4$\times 10^{-5}$   & -4.92735 $\pm$ 0.00059 \\
59910.6317 & -16487.864 $\pm$ 0.528 & 7380.76 $\pm$ 1.06   & -41.41 $\pm$ 1.06     & 52.9909 $\pm$ 0.0076     & 0.163737 $\pm$ 0.000115 & 0.198409 $\pm$ 4.4$\times 10^{-5}$ & -4.92001 $\pm$ 0.00065 \\
59913.7050 & -16482.505 $\pm$ 0.324 & 7388.19 $\pm$ 0.65   & -40.49 $\pm$ 0.65     & 52.8724   $\pm$ 0.0046      & 0.175557 $\pm$ 4.5$\times 10^{-5}$  & 0.202343 $\pm$ 2.5$\times 10^{-5}$ & -4.85833 $\pm$ 0.00022 \\
59915.6274 & -16486.882 $\pm$ 0.612 & 7388.31 $\pm$ 1.22   & -36.27 $\pm$ 1.22     & 52.9597 $\pm$ 0.0088      & 0.166462 $\pm$ 0.000138 & 0.200375  $\pm$ 5.5$\times 10^{-5}$ & -4.90500 $\pm$ 0.00075 \\
59917.7656   & -16489.050 $\pm$ 0.355 & 7384.57 $\pm$ 0.71  & -41.46 $\pm$ 0.71     & 52.9032 $\pm$ 0.0051      & 0.172896 $\pm$ 5.4$\times 10^{-5}$  & 0.20123 $\pm$ 2.7$\times 10^{-5}$ & -4.87147 $\pm$ 0.00027 \\
59919.5689 & -16484.671 $\pm$ 0.470 & 7383.81 $\pm$ 0.94   & -41.445 $\pm$ 0.94     & 52.9498  $\pm$ 0.0067     & 0.16945 $\pm$ 9$\times 10^{-5}$    & 0.199559 $\pm$ 3.9$\times 10^{-5}$ & -4.88911 $\pm$ 0.00047 \\
\enddata
\label{espresso_vals}
\end{deluxetable}

\section{\texttt{allesfitter} Corner Plots}
\restartappendixnumbering
In this appendix, we show the corner plots from our \texttt{allesfitter} fit.

\begin{figure*}[!h]
    \centering
    \includegraphics[width=\textwidth]{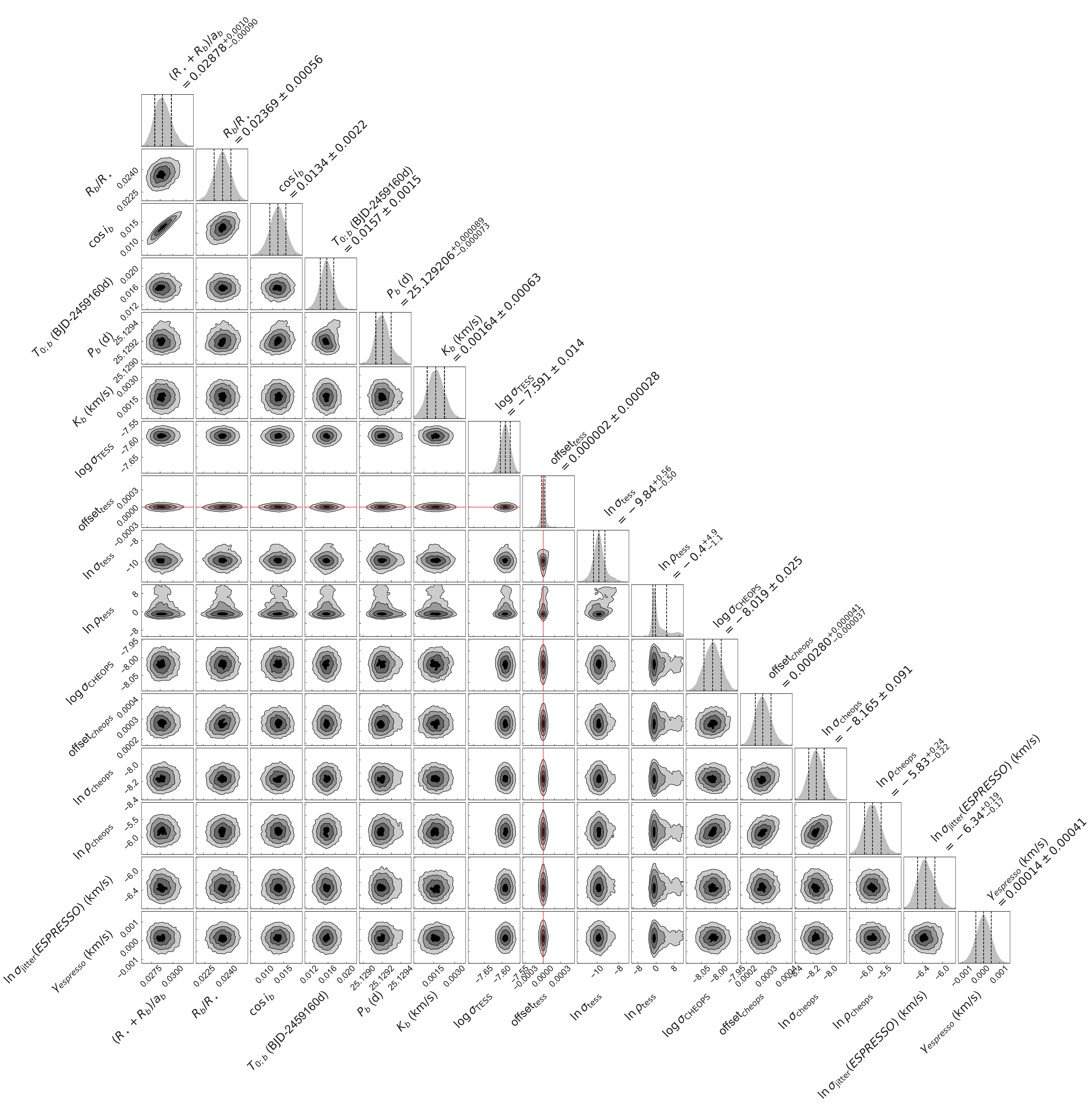}
    \caption{Corner plots of modeled parameters obtained from \texttt{allesfitter}.}
    \label{fig:corner_fitted}
\end{figure*}

\begin{figure*}[!h]
    \centering
    \includegraphics[width=\textwidth]{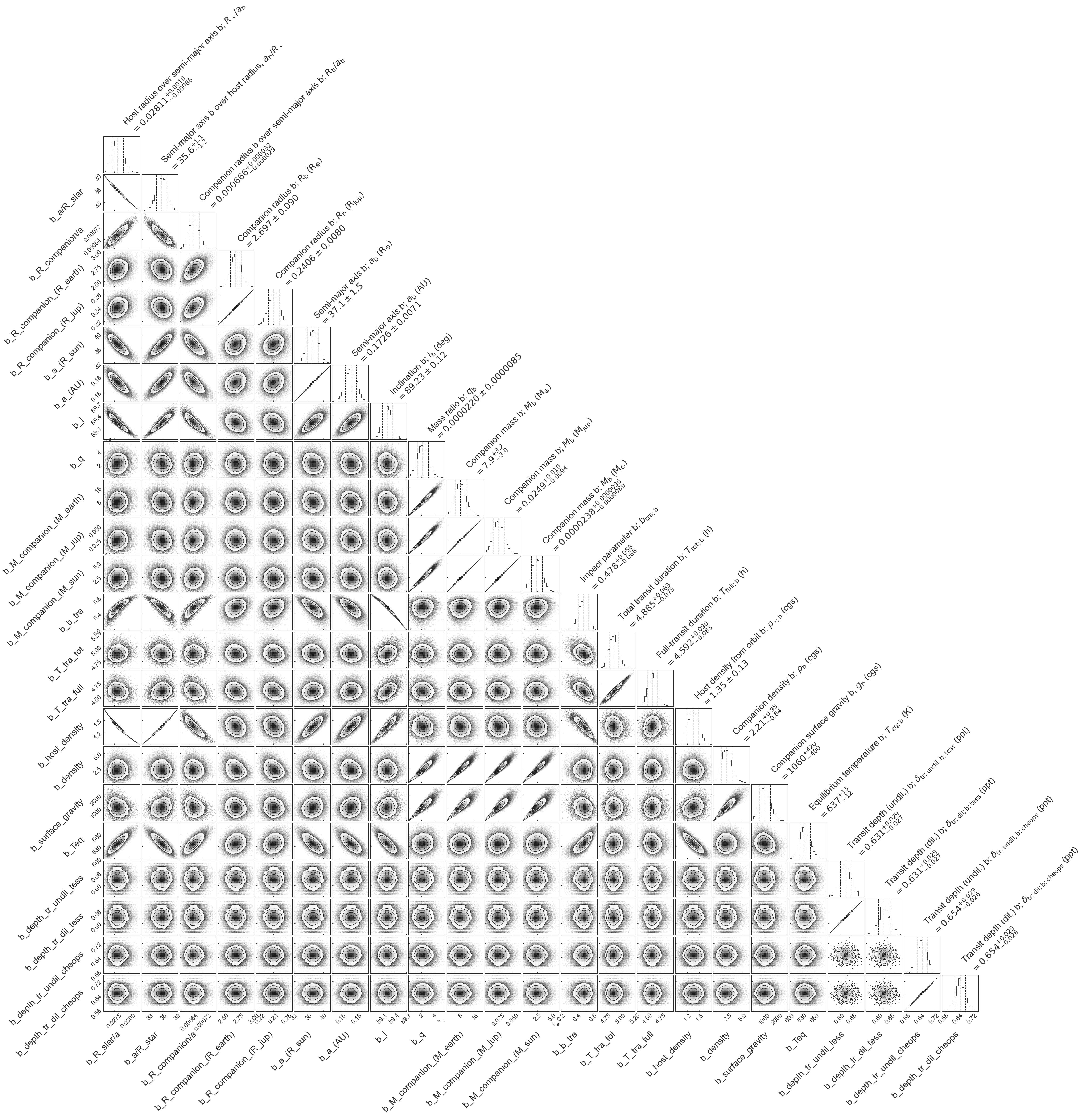}
    \caption{Corner plots of derived parameters obtained from \texttt{allesfitter}.}
    \label{fig:corner_derived}
\end{figure*}
%\end{appendix}

\end{document}